\documentclass[manuscript,trackchanges]{aastex62}

\usepackage{amsmath} 

\graphicspath{{./}{Figures/}}

\received{16 Aug, 2019}
\revised{xx xx, 2019}
\accepted{4 Jan 2020}

\submitjournal{ApJ}

\begin{document}

\title{ALMA observations of two massive and dense MALT90 clumps}

\correspondingauthor{S., Neupane}
\email{sneupane@das.uchile.cl}

\author[0000-0001-9596-9938]{Sudeep Neupane}
\affiliation{Departamento de  Astronom{\'i}a, Universidad de Chile, 
Camino el Observatorio 1515,
Las Condes, Santiago, Chile}

\author{Guido Garay}
\affiliation{Departamento de  Astronom{\'i}a, Universidad de Chile, 
Camino el Observatorio 1515,
Las Condes, Santiago, Chile}

\author{Yanett Contreras}
\affiliation{Leiden Observatory, Leiden University, PO Box 9513, NL-2300 RA Leiden, the Netherlands}

\author{Andres Guzm{\'a}n}
\affiliation{NAOJ, Chile Observatory, East Asia ALMA Regional Center, Mitaka, Tokyo 181-8588, Japan}

\author{Luis Felipe Rodr{\'\i}guez}
\affiliation{Instituto de Radioastronom{\'i}a y Astrof{\'i}sica, Universidad Nacional Aut{\'o}noma de M\'exico, 
Apdo. Postal 3-72 (Xangari), 58090 Morelia, Michoac{\'a}n, M\'exico}

\begin{abstract}

We report Atacama Large Millimeter Array observations of 3 mm dust continuum emission and line emission, in HCO$^{+}$, H$^{13}$CO$^{+}$, N$_{2}$H$^{+}$ and CH$_{3}$CN, towards two massive and dense clumps (MDCs) in early but distinct evolutionary phases (prestellar and
protostellar), made with the goal of investigating their fragmentation characteristics at angular scales of
$\sim$1$\arcsec$. Towards the prestellar clump we detected ten compact structures (cores), with radius
from 1200 to 4500 AU and masses from 1.6 to 20~M$_\odot$. Half of these cores exhibit inverse P Cygni
profiles in HCO$^{+}$ and are subvirialized indicating that they are undergoing collapse. Towards the
protostellar clump we detected a massive (119~M$_\odot$) central core, with a strong mass infall rate, and
nine less massive cores, with masses from 1.7 to 27~M$_\odot$ and radius from 1000 to 4300 AU.
CH$_{3}$CN rotational temperatures were derived for 8 cores in the protostellar clump and 3 cores in the
prestellar clump.
Cores within the prestellar clump have smaller linewidths and lower temperatures than cores within the
protostellar clump. The fraction of total mass in cores to clump mass is smaller in the prestellar clump
($\sim$6\%) than in the protostellar clump ($\sim$23\%). We conclude that we are witnessing the evolution
of the dense gas in globally collapsing MDCs; the prestellar clump illustrating the initial stage of
fragmentation, harboring cores that are individually collapsing, and the protostellar clump reflecting a later
stage in which a considerable fraction of the gas has been gravitationally focused into the central region.

\end{abstract}


\section{Introduction} \label{sec:intro}

A wealth of observations have shown that filamentary structures are ubiquitous within molecular clouds (eg., \citealt{schneider-1979, Myers2009, hershel, Andre-2010}). These long molecular structures are inhomogeneous and present over-densities, most likely a result of fragmentation (eg., \citealt{Takahashi2013, teixeira2016, Contreras-2016}).  It is in the most massive ($\sim$10$^{3}$ M$_{\odot}$) and dense ($\sim$10$^{4}$~cm$^{-3}$) overdensities, which we refer as  massive and dense clumps (or MDCs), where high-mass stars form (\citealt{faundez-2004,contreras2017}). However, the early evolution of MDCs  and the ensuing fragmentation leading to the formation of  cores is not well understood. The relative importance of primordial clump fragmentation versus large-scale accretion in determining the distribution of core masses still remains to be assessed. Recent ALMA observations with moderate angular resolution ($\sim$3.5$\arcsec$) towards a sample of MDCs in early evolutionary stages (infrared quiet)  with masses in the range from 200 to 2000 M$_{\odot}$ revealed limited fragmentation at the scale of $\sim$0.1 pc, showing typically $\sim$3 cores, and a high efficiency in the formation of high-mass cores (\citealt{Csengeri2017}).  Observations with higher resolution  ($\sim$0.03-0.05 pc) of clumps with similar characteristics show a range of substructures -- from a few fragments (eg., \citealt{Peretto2013, Sanhueza2017}) to ten or more fragments (eg., \citealt{Lu2018}, \citealt{Contreras2018}). Some works concluded that the fragmentation properties of clumps are described by gravo-turbulence (eg., \citealt{Zhang2015}) while others find them consistent with pure thermal Jeans fragmentation (eg., \citealt{Palau2015, Palau2018}). 
\cite{teixeira2016} found that the separation of clumps within a filamentary cloud is consistent with the Jeans length of the filament 
while the separation between the individual cores within the clumps is smaller than the Jeans length of the clump, which they suggest indicates that the local collapse of the clumps ocurrs at a much faster pace than the global collapse of the filament.   
  
Determining the physical and kinematical properties of the molecular gas in MDCs at both, the large clump scale ($\sim 1$ pc) and small core scale (5000 AU), will permit to investigate the presence of global or localized collapse and the characteristics of the primordial fragmentation. These properties together constitute a key discriminator between current models of the fragmentation and evolution of MDCs, 
such as Competitive accretion (\citealt{bonnel-2006}), Turbulent fragmentation  \citep{Padoan2002}, Hierarchical gravitational fragmentation (\citealt{Vazquez2009, Vazquez2017}).

In this work we present a study of two MDCs, one in the prestellar stage and the other in the protostellar stage of evolution, using high resolution ALMA Band 3 continuum and molecular line observations with the goal of identifying and determining the physical characteristics of the dense and compact structures within the MDCs and to test models of the fragmentation and 
evolution of MDCs, and possibly to guide future theories. In \S \ref{sec:s2} we briefly review the characteristics of the observed MDCs. In \S \ref{sec:s3} and  \S \ref{sec:s4} we describe the observations and present the results, respectively.  In \S \ref{sec:s5} we discuss the  analysis of the continuum and molecular observations. In \S \ref{sec:s6} we compare our results with the predictions of different models. 

\begin{figure*}[ht!]
\centering
\includegraphics[trim={1cm 1cm 1cm 1cm}, clip, width=0.49\linewidth]{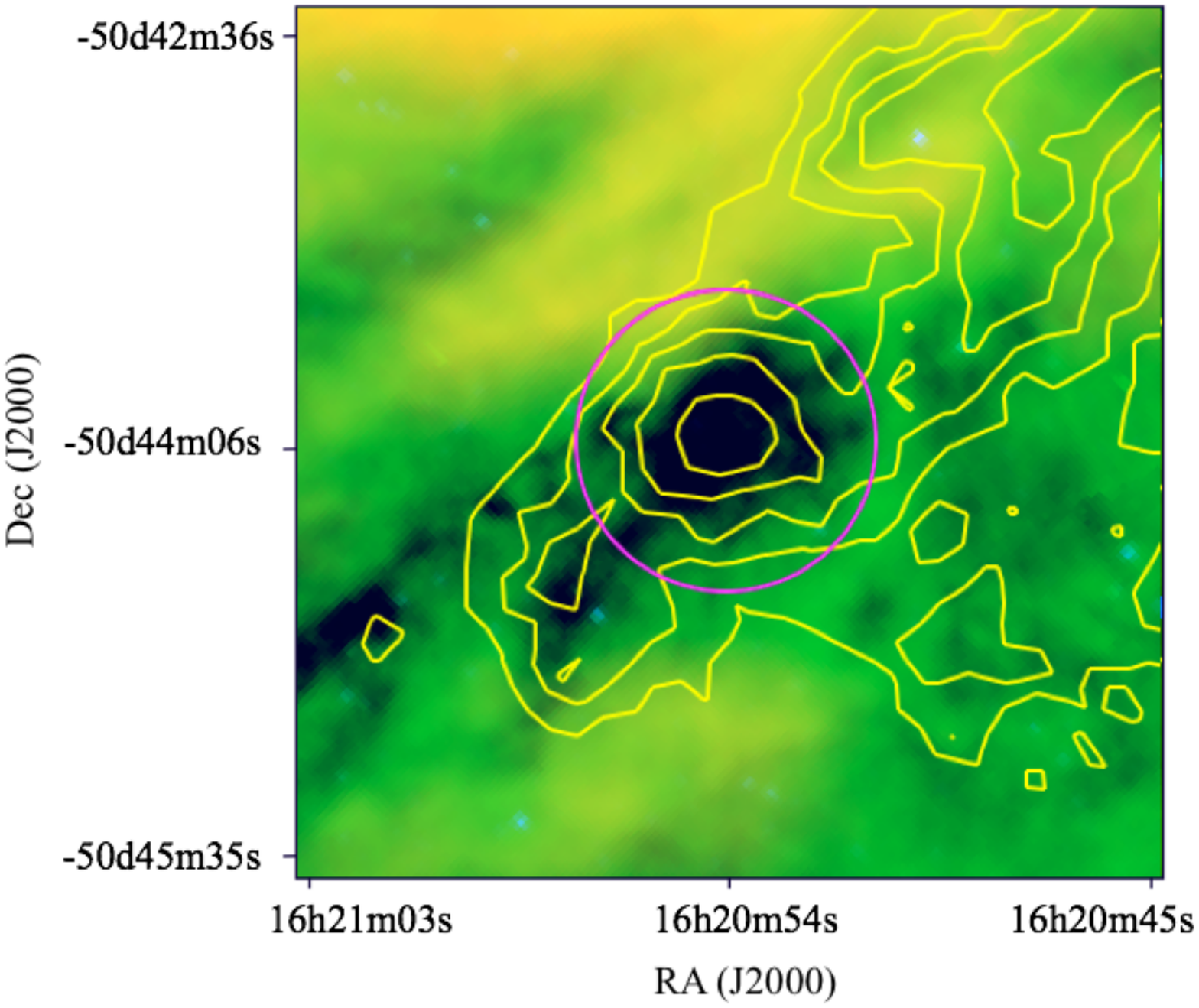}
\includegraphics[trim={1cm 1cm 1cm 1cm}, clip, width=0.49\linewidth]{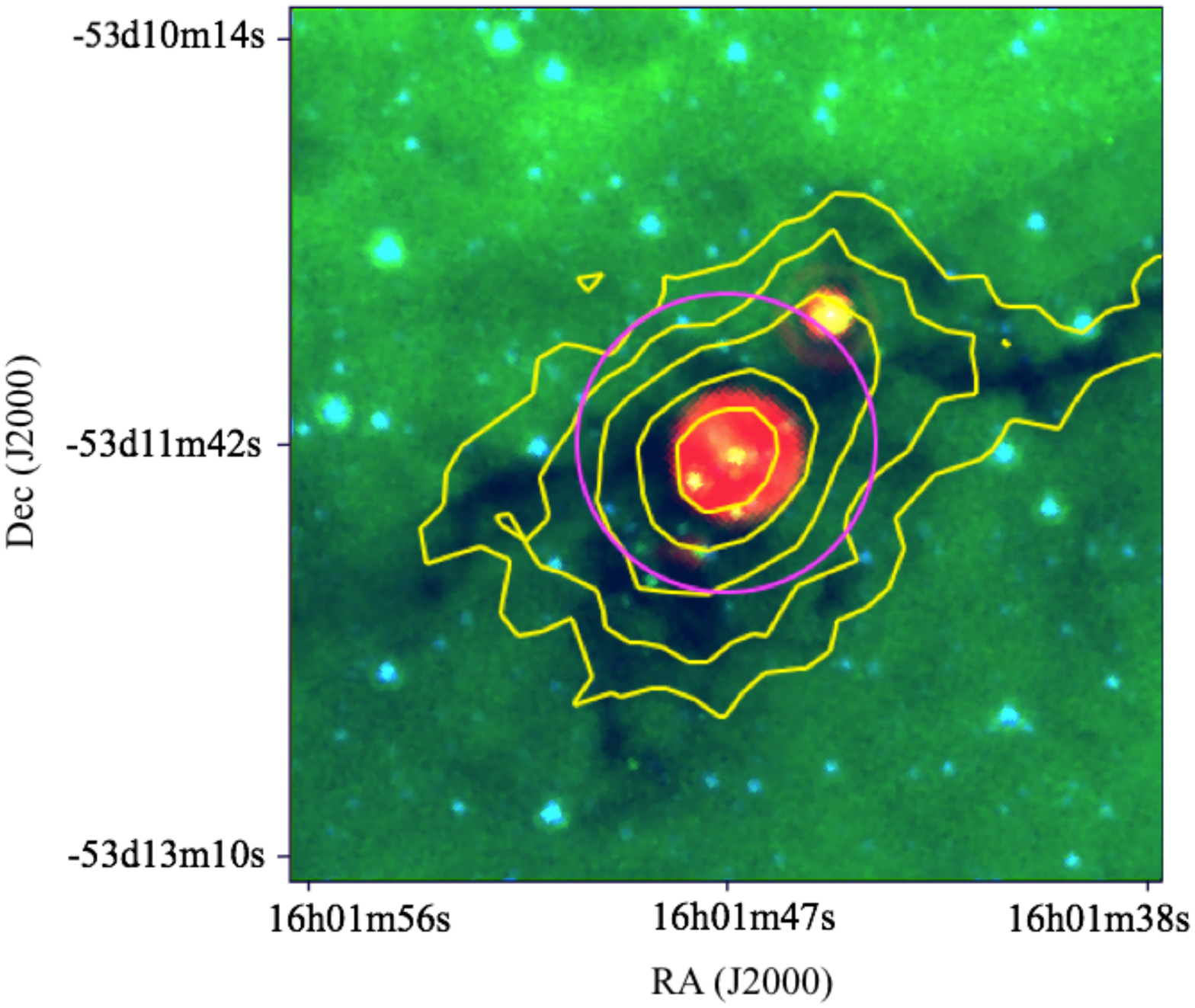}
\caption{Three colour Spitzer images (blue: 3.6 $\mu$m, green: 8 $\mu$m and red: 24 $\mu$m)(IRAC: \citealt{SpitzerIRAC}, MIPS: \citealt{SpitzerMIPS}) of the MALT90 targets overlaid with contours of the 870 $\mu$m emission from the Atlasgal survey (FWHM$\sim$20$\arcsec$).  The magenta circle indicate the ALMA primary beam (FWHM) of $\sim$62$\arcsec$ in Band 3.
Left panel: Prestellar clump AGAL333.  Contour levels are drawn at 3$\sigma$, 5$\sigma$, 7$\sigma$, 9$\sigma$
and 11$\sigma$  ($\sigma$ = 93.3 mJy beam$^{-1}$). Right panel:  Protostellar clump AGAL329.  Contour levels are drawn at 3$\sigma$, 6$\sigma$, 12$\sigma$, 24$\sigma$ and 48$\sigma$ ($\sigma$ = 89.3 mJy beam$^{-1}$).  \label{fig:f1} }
\end{figure*}

\section{The targets \label{sec:s2}}

The two MDCs studied in this work were selected from the MALT90\footnote{The Millimetre Astronomy Legacy Team 90 GHz Survey: http://malt90.bu.edu} catalog (\citealt{Rathborne-2016}), one AGAL333.014-0.521 (hereafter AGAL333) classified as been in the prestellar stage and the other AGAL329.184-0.314 (hereafter AGAL329) classified in the protostellar stage of evolution (see Figure \ref{fig:f1}). The MALT90 project (\citealt{jackson-2013,malt902}) surveyed, with the MOPRA telescope, the emission in 15 different molecular lines (mostly J=1$\rightarrow$0 transitions) and one recombination line towards $\sim$3200 MDCs.  Towards AGAL329 and AGAL333 emission was detected in, respectively,  9 and 8 lines,  including the high density tracers HCO$^{+}$, HNC, HCN and N$_{2}$H$^{+}$. The J=1$\rightarrow$0
transitions of these four species have critical densities\footnote{Defined as, n$_{crit}$ = $A_{ul}$/$\gamma_{ul}$, where $A_{ul}$ is Einstein coefficient and $\gamma_{ul}$ is the collisional rate.} of the order of 10$^{5}$-10$^{6}$ cm$^{-3}$ for a temperature of 20 K, indicating that the clumps indeed have high densities, a requisite for the formation of high-mass stars. The spectra of the optically thick HCO$^{+}$ emission from the protostellar clump shows a double peak profile, with the blueshifted peak being stronger than the redshifted peak, while the spectra of the optically thin H$^{13}$CO$^{+}$ emission shows a single line with a peak velocity in between the velocities of the blue and red peaks of the HCO$^{+}$ line. These profile characteristics are signposts of infall motions (e.g, \citealt{Anglada1987, mardones1997, Devries2005}) suggesting that AGAL329 is undergoing a large scale collapse. On the other hand, the profiles of the HCO$^{+}$ and H$^{13}$CO$^{+}$ emission from the prestellar clump are nearly Gaussian, indicating a more static, quiescent region. As expected, SiO emission, which traces outflow/shocked gas (eg., \citealt{Martin-Pintado1992}), is only detected towards AGAL329. 

\begin{table*}
\begin{center}
\caption{Observed and derived parameters of the clumps.\label{sampleparams}} 
\begin{tabular}{lccc}
\hline \hline
    \multicolumn{1}{l}{Parameter}  & 
    \multicolumn{1}{c}{AGAL333.014-0.521} &
    \multicolumn{1}{c}{AGAL329.184-0.314}  &  \multicolumn{1}{c}{Reference} \\ \hline 
Clump type					& Prestellar	& Protostellar	& (1)  \\
Distance (kpc) 				& 3.72		&	3.46 	&	(2)\\
V$_{lsr}$ (km s$^{-1}$)			& -53.8		&    -50.5  	&	(1) \\
Line width  (km s$^{-1}$)			& 3.1			&	3.4	&	(1) \\
$F_{870\mu m}$ (Jy)			& 16.42		&	23.02	 & (3),(4)  \\
$\theta$ ($\arcsec\times\arcsec$) & 43$\times$17	&	24$\times$15	&  (3),(4) \\ 
Size   (pc) 				&0.49		&	0.32	     &	(5)  \\
T$_{dust}$ (K)				& 22			&	28		&	(5) \\
M$_{dust} (M_{\odot}$)		&1080		&	940		&   (5)   	\\

n(H$_{2}$) (10$^{5}$cm$^{-3}$)	&0.32		&	1.00	   &  (5) 	\\ % mu = 2.8 used
M$_{vir} (M_{\odot}$)  		& 980		&	770		&	\\
$\alpha_{vir}$ 				& 0.91		&	0.82		&	\\
Jeans mass  (M$_{\odot}$)		& 6.8			&	5.5		&	\\ 
Jeans radius  (pc)      			& 0.09       	&  	0.06  &  \\ \hline

\end{tabular}
\end{center}
\tablerefs{ (1) \citealt{Rathborne-2016}; (2) \citealt{Whitaker2017}; (3) \citealt{Contreras-2013a}; (4) \citealt{Urquhart-2014}; (5) this work.}
\end{table*}

\begin{figure*}
\plottwo{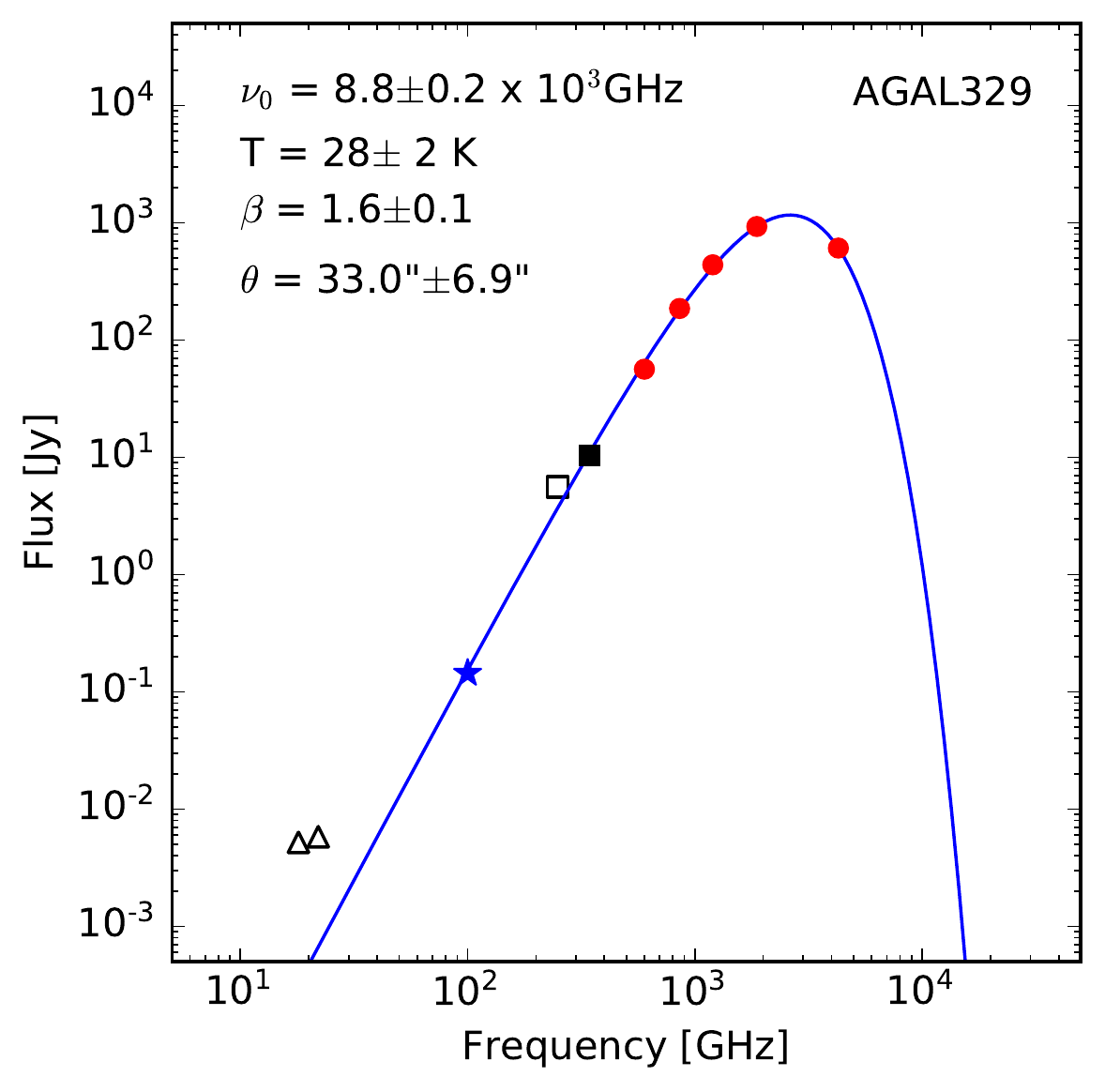}{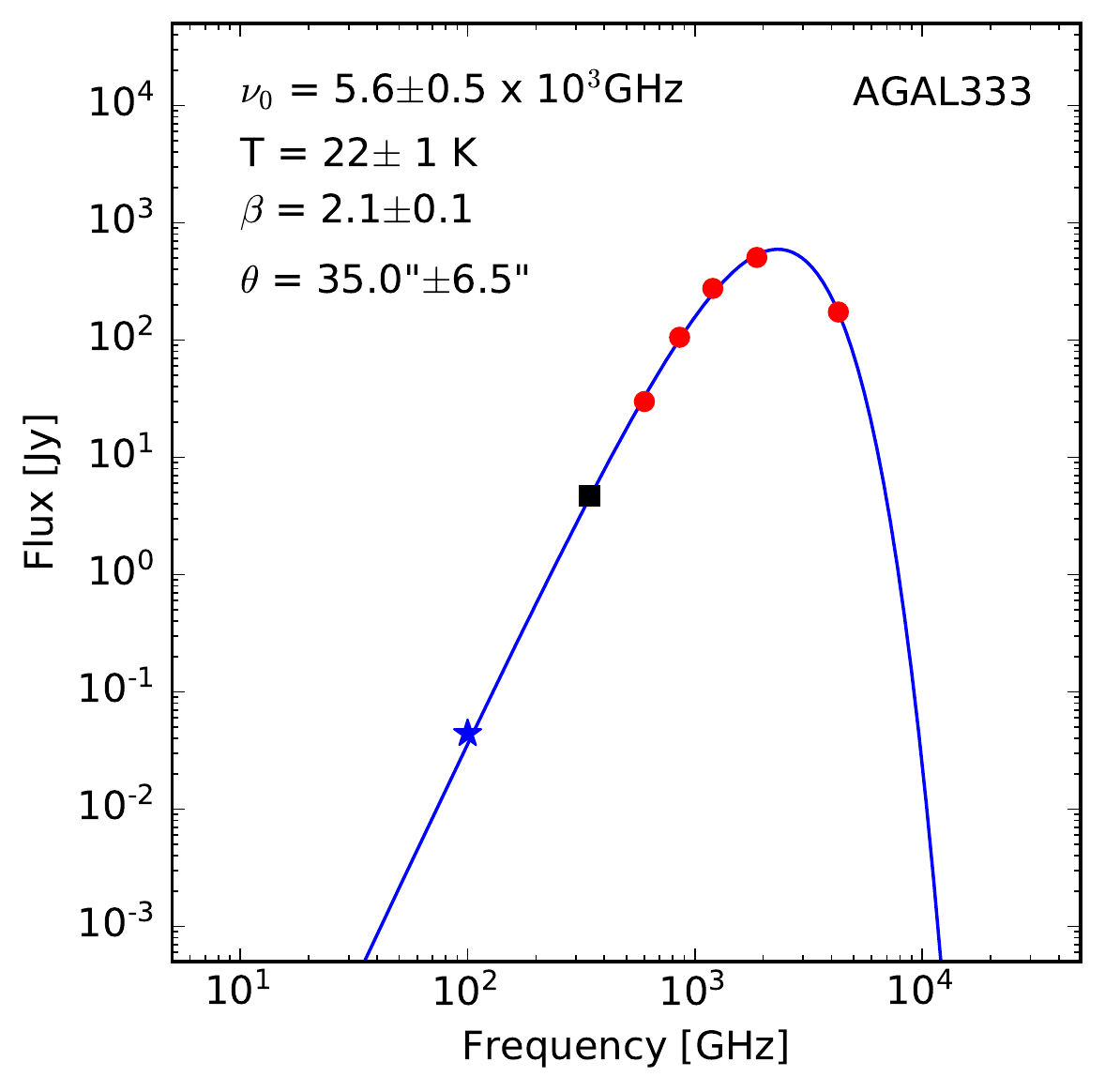} 
\caption{Spectral energy distribution of clumps AGAL329  (left) and AGAL333 (right). Symbols are described in the text.  The continuous lines (blue) correspond to the SEDs fit obtained from a least squares method. The fitted parameters are given inside each box. \label{fig:Seds}}
\end{figure*}

Table \ref{sampleparams} lists observed and derived parameters of the clumps. 
The velocities and line widths correspond to those determined from the hyperfine fitting of the N$_{2}$H$^{+}$ emission as observed with the MOPRA telescope in the MALT90 survey.
The dust temperatures were determined by us from a fit to the spectral energy distributions (SEDs), shown in Figure \ref{fig:Seds},
using a single temperature modified blackbody model  (c.f.,  \citealt{Guzman2015, konig2017}) , 
\begin{eqnarray}\label{eq:1}
\begin{array}{l}
S_{\nu} = \Omega_s B_{\nu}(T_d)(1-e^{-\tau_{\nu}})  ~~,
\end{array}
\end{eqnarray}
where S$_{\nu}$,  B$_{\nu}$,  T$_d$ and $\Omega_s$ are, respectively, the flux density, Planck function, dust temperature
and effective solid angle subtended by the clump. We assume that the dependence of the optical depth, $\tau_{\nu}$, with frequency $\nu$ can be expressed as
\begin{eqnarray}
\tau = (\nu / \nu_{o})^{\beta} ~~,
\end{eqnarray}
where $\nu_{o}$ is the frequency at which the dust opacity is unity and $\beta$ is the spectral index of the dust absorption coefficient. 

The data points at infrared wavelengths (70 to 500 $\mu$m; red circles) were obtained from the Hi-Gal images (\citealt{hershel}) available in the  Herschel Science Archive and the data point at 850 $\mu$m (black square) was obtained from ATLASGAL images (\citealt{Schuller}). The fluxes were extracted using simple aperture photometry of a circular region with radii of 27$\arcsec$ and 20$\arcsec$ for the prestellar and protostellar clumps, respectively. Errors in the flux densities, mostly due to calibration uncertainties, are less than 30\%. Error bars are then smaller than the size of the symbols. Also incorporated in the SED are the flux densities measured at 100 GHz  using the ACA array alone (present work; blue stars) and for the protostellar clump the 1.2 mm flux density reported by \citealt{Beltran2006} (open square). A least squares fit to the SED, using the SciPy optimize module in Python (\citealt{scipy2019}),
gave values of T$_d$ and $\beta$ of 22 K and 2.1 for the prestellar clump and 28 K  and 1.6 for the protostellar clump.  Shown in the SED of the  protostellar clump (Figure \ref{fig:Seds} - left panel) are the flux densities at 18 GHz and 22 GHz observed with ATCA (\citealt{sanchez-monge2013}; triangles).  These flux densities are well above those expected from the dust emission model,  and most likely  correspond to free-free emission from either an UC HII region or a region of  shocked gas. 

The clump size, mass, and density, given in lines 7, 9 and 10  in Table \ref{sampleparams}, were derived from the 870 $\mu$m continuum emission.  
The last two parameters were computed using  the dust temperatures derived from the SED, a dust absorption coefficient of 1.85 cm$^{2}$
gr$^{-1}$ (\citealt{Ossenkopf1994}), and a gas to dust ratio of 100. The virial masses,  $M_{vir}$ = 5$\sigma^{2} R/$G,  where $\sigma$ = $\Delta$V/(8ln2)$^{1/2}$, are 980 
M$_{\odot}$ for the prestellar clump and 770 M$_{\odot}$ for the protostellar clump. The virial parameter, defined as  $\alpha_{vir}$ = $M_{vir}/M_{dust}$, is 0.91  for the prestellar clump and 0.82 for the protostellar clump,  suggesting that both of them are gravitationally bound. Also given in Table \ref{sampleparams} are the Jeans mass and Jeans radius at the average temperature and density for both clumps.

In summary, the clumps selected for this study, which are at similar distances fom the Sun, have masses, sizes and
densities characteristics of high-mass star forming regions and their virial parameters indicate that they are gravitationally bound.  The IR and molecular line observations suggest, however, that they are in different evolutionary stages.  AGAL329 harbors a strong 24 $\mu$m point source and show line profiles characteristics of infalling motions, indicating it is in a more advanced stage of evolution than AGAL333, which exhibits Gaussian profiles and no energy sources at IR wavelengths.

\section{Observations \label{sec:s3}}

\begin{table}
\centering
\caption{Observational parameters. \label{spectralsetup}}
\begin{tabular}{lcccdd}
\hline \hline
SPW		& Center	Freq.		& Bandwidth 	& Vel. res. 	& \multicolumn{2}{h}{On source integration time} \\
		& (MHz)		&	(MHz)		& (km s$^{-1}$)				&	(12m	 Array)	&	(ACA)			\\ \hline
\multicolumn{6}{c}{Observing setup I } \\ \hline
N$_{2}$H$^{+}$ 	&93173.402			&	117.19		&	0.393	&	23 min		&	32 min			\\
CH$_{3}$CN		&91985.284			&	117.19		&	0.398	&			&			\\
Cont. 1			&92500.000			&	1875.00		&	101.262	&			&			\\
Cont. 2			&103500.000			&	1875.00		&	90.500	&			&			\\
Cont. 3			&105400.000			&	1875.00		&	88.869	&			&			\\ \hline
\multicolumn{6}{c}{Observing setup II} \\ \hline
HCO$^{+}$		& 89188.526			&	117.19		&	0.205 	&	23 min		&	32 min		\\
H$^{13}$CO$^{+}$	& 86754.288			&	117.19		&	0.211	&			&			\\
Cont. 4			& 99000.000			&	1875.00		&	94.614	&			&			\\
Cont. 5			& 100900.000			&	1875.00		&	92.832	&			&			\\ \hline
\end{tabular}
\end{table}

We made ALMA Band 3 (86-116 GHz) dust continuum and molecular line observations towards the MDCs AGAL333 and AGAL329.  Since the sizes of these MDCs are smaller than the ALMA field of view at 3 mm ($\sim$62$\arcsec$), single pointing observations were carried out, as part of Cycle 4, during Dec 2016 and Jan 2017 using both  the 12 m array and 7m Atacama Compact Array (ACA). We used two different spectral set ups (see Table \ref{spectralsetup}). In the first one, five separate spectral windows (SPW) were used, three for continuum observations and two for observations of the N$_{2}$H$^{+}$J=1$\rightarrow$0 and 
CH$_{3}$CN J=5$\rightarrow$4 lines. In the second set up, the bandwidth was separated into four SPWs, two for continuum observations and two for observations of the HCO$^{+}$J=1$\rightarrow$0  and H$^{13}$CO$^{+}$ J=1$\rightarrow$0 lines.  The observed molecular species were chosen for the following reasons. N$_{2}$H$^{+}$ suffers little depletion and is one of the best tracers of the dense and cold gas (\citealt{Bergin1997,Caselli-2002b}). The emission in the HCO$^{+}$ and H$^{13}$CO$^{+}$ lines are, respectively, usually optically thick and thin towards dense clumps and therefore their simultaneous observations are useful to probe the presence of infall or expansion motions. The CH$_{3}$CN molecule is a good temperature probe, of both the  large scale diffuse gas and small scale dense gas (eg., \citealt{Guesten1985,blake-1987, Araya2005,Jones2008}).

\begin{table*}[htbp!]
\centering
\caption{Synthesized beam and rms noise from 7m+12m combined maps.  \label{obsparams}}
\begin{tabular}{lrcccccccc}\hline \hline
Maps		&	& \multicolumn{3}{c}{AGAL329} 	& \multicolumn{1}{c}{}& & \multicolumn{3}{c}{AGAL333}		\\  \cline{2-5} \cline{8-10}
			&	&  rms noise		& 	Beam  & PA	& & &  rms noise	& Beam	& PA  \\ 
			&	&  (mJy/beam) 		& ($\arcsec\times\arcsec$) & ($\degr$)	& & &  (mJy/beam) 	& ($\arcsec\times\arcsec$) &  ($\degr$) \\  \hline

Continuum	&	& 8.5$\times 10^{-2}$ & 1.5 x 1.3 & 155 & 	& & 4.7$\times 10^{-2}$ & 1.5 x 1.3 & 138 \\
HCO$^{+}$	&	&	4.0			&	1.6 x 1.3 &  155	& & &	4.0				&	1.6 x 1.3 & 147	\\
H$^{13}$CO$^{+}$&	&	3.7			&	1.7 x 1.3 &  155	& & &	3.7				&	1.7 x 1.3 & 146	\\
N$_{2}$H$^{+}$&	&	4.0			&	2.1 x 1.7 & 66	& & &	4.0				&	2.1 x 1.7 & 67	\\
CH$_{3}$CN	&	&	3.9			&	2.1 x 1.8  & 68	& & &	3.7				&	2.1 x 1.8 & 68	\\ \hline
\end{tabular}
\end{table*}

The integration time on source in each of the setups was 23  and 32 minutes for the  7m and 12m array observations,
respectively. The bandpass and flux calibrations were carried out using multiple quasars (J1603$-$4904, J1617$-$5848, J1427$-$4206, J1603$-$4904, J1312$-$0424, J1617$-$5848, J1650$-$5044, J1924$-$2914, J2131$-$1207), Mars and Callisto. 
Data calibration and reduction were made using the Common Astronomy Software Application (CASA: \citealt{casa2007}) version 4.7 package. Independently calibrated 12m and 7m dataset were concatenated and cleaned together using the CASA {\it tclean} task with a  {Briggs} weighting of 0.5. We used a {\it multi-scale} clean deconvolver (\citealt{cornwell2008}), with scale values of 0, 6, 10 and 30 times the image pixel size (0.3$\arcsec$). For the continuum imaging we concatenated all 5 continuum spectral windows. We used interactive mode for continuum imaging while spectral cubes were made using continuum subtracted spectra with automated masking procedure {\it auto-multithresh} using noise threshold parameter {\it noisethreshold} of 2 sigma. This parameter corresponds 
to the minimum signal-to-noise value that is masked. This technique mimics what we would do in manual masking in interactive cleaning procedure. From the final spectral line cubes, integrated intensity and velocity maps were made using casa task {\it immoments}. 
The angular resolution achieved in the continuum observations are 1.46$\arcsec\times1.34\arcsec$ (P.A. 155$\degr$) and 1.46$\arcsec$x1.33$\arcsec$  (P.A. 138$\degr$)   for AGAL329 and AGAL333, respectively. The rms values achieved in the continuum and line maps are listed in the Table \ref{obsparams}. 
For the molecular lines the rms reported is that determined from line-free channels.

\section{Results \label{sec:s4}}

\subsection{Continuum emission}
\begin{figure*}[ht!]
\centering
\includegraphics[trim={0.3cm 0.21cm 0.35cm 0.6cm}, clip, width=0.61\textwidth]{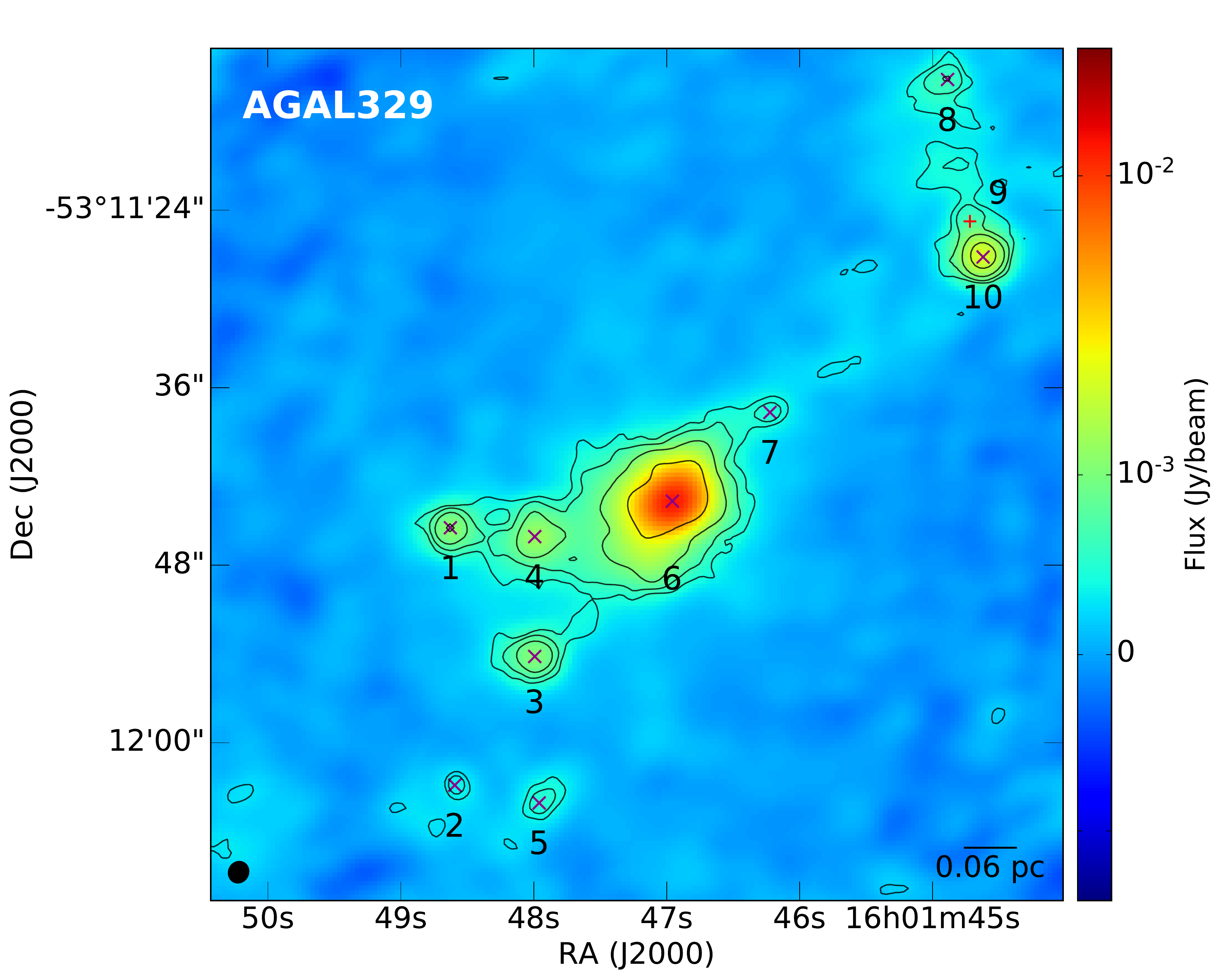}\\
\includegraphics[trim={0.3cm 0.21cm 0.35cm 0.6cm}, clip, width=0.61\textwidth]{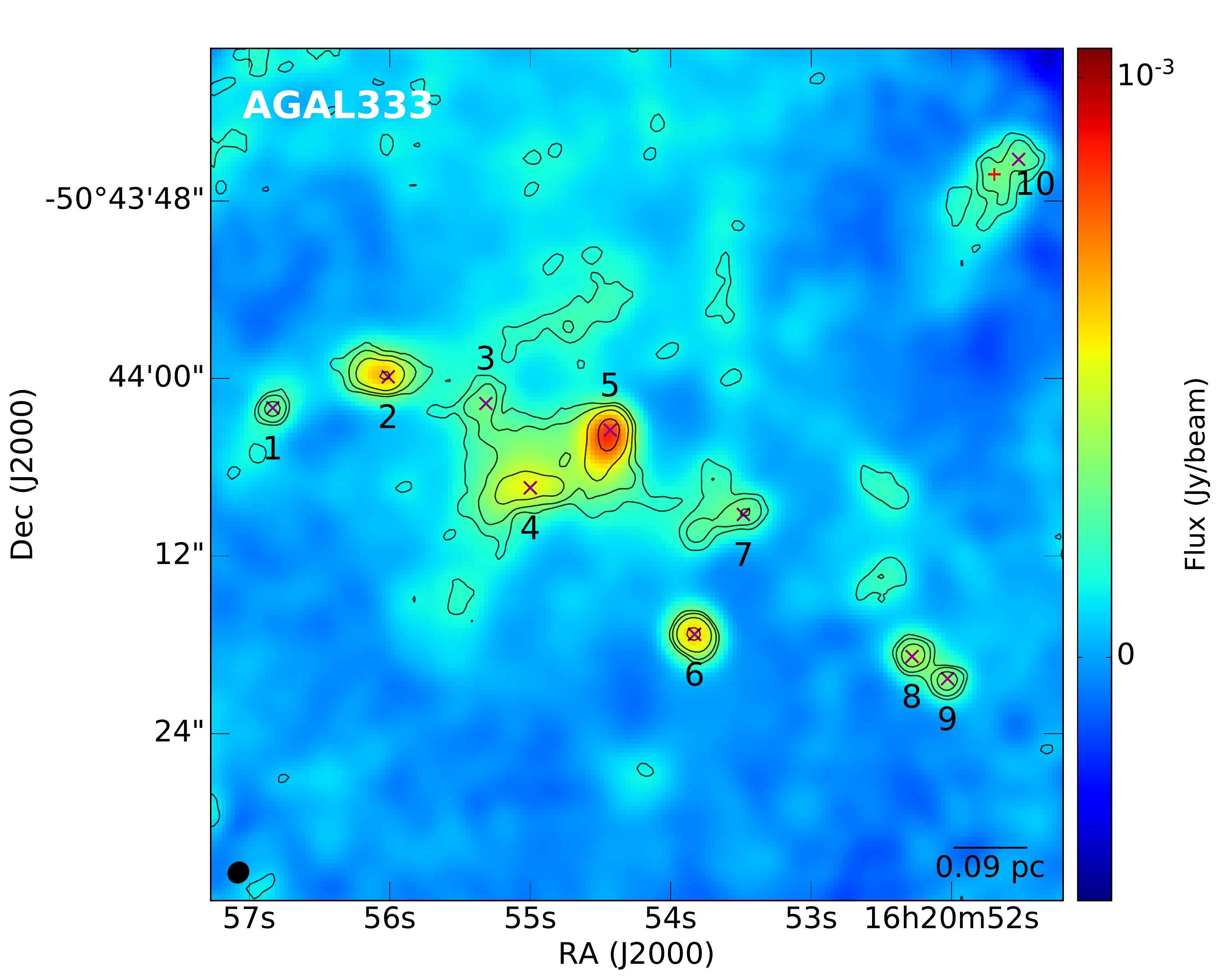}
\caption{Images of the 3 mm continuum emission observed with ALMA towards the protostellar clump AGAL329 (upper panel) and prestellar clump AGAL333 (lower panel). The magenta crosses indicate the cores extracted in common by Astrodendro  and Clumpfind. The `+' symbols in red indicate cores identified only by Clumpfind or Astrodendro. The scale bar in the bottom right corner indicates the Jeans radius of the clumps. Also shown are contour levels of the 3mm emission, drawn at  3, 5, 10 and 30 $\sigma$, where $\sigma$ are 4.7$\times10^{-2}$ and  8.5$\times10^{-2}$ mJy/beam for AGAL333 and AGAL329, respectively. The black ellipse shown at the bottom left corner indicates the beam size.}
\label{fig:almaimages}
\end{figure*}

Figure \ref{fig:almaimages} shows our ALMA images of the 3 mm continuum emission towards AGAL329 and AGAL333. The emission from AGAL329 arises from an extended, bright central source and a handful of compact, weaker structures. The emission from AGAL333 arises from several compact structures spread out across the region, most of them being aligned in a NE-SW direction.

\begin{table*}[h!]
\fontsize{8.}{8.}\selectfont
\caption{Observed parameters  of the cores.}
\begin{center}
\footnotesize
\begin{tabular}{cccccccccd}
    \hline \hline
      Core &    \multicolumn{2}{c}{Peak position} &
    \multicolumn{3}{c}{Clumpfind} &
    \multicolumn{3}{c}{Astrodendro} &
    \multicolumn{1}{h}{Remarks} \\  \cline{2-3} \cline{5-6} \cline{8-9}
 	&	RA	&	DEC	&	&	Flux	&	Ang.	size	&	& Flux	&	Ang. 	size		&			\\
       &	hh:mm:ss	&	dd:mm:ss		&		& (mJy)	&		(HWHM $\arcsec$)	&	&	(mJy)	&		(HWHM $\arcsec$)	&	      \\	 \hline 
     \multicolumn{10}{c}{AGAL329} \\ \hline 
mm-1	&	16:01:48.62  &	-53:11:45.47	&	&4.69	&	0.65	&				&	3.78	&	0.65	&							\\
mm-2	&	16:01:48.59  &	-53:12:02.87 	&	&0.49	&	0.29	&				&	0.50	&	0.41	&							\\
mm-3	&	16:01:47.99  &	-53:11:54.17	&	&5.12	&	0.75	&				&	5.17	&	1.09	&							\\
mm-4	&	16:01:47.99  &	-53:11:46.07	&	&9.84	&	1.14	&				&	5.60	&	0.84	&							\\
mm-5	&	16:01:47.95  &	-53:12:04.07 	&	&1.04	&	0.50	&				&	1.05	&	0.61	&							\\
mm-6	&	16:01:46.95 & 	-53:11:43.67	&	&102.15	&	1.77	&				&	93.55	&	1.66	&							\\
mm-7	&	16:01:46.22 &	-53:11:37.67 	&	&1.00	&	0.51	&				&	0.61	&	0.45	&							\\
mm-8	&	16:01:44.88 &	-53:11:15.17 	&	&2.48	&	0.77	&				&	2.50	&	1.06	&							\\
mm-9\tablenotemark{a}	&	16:01:44.72 &	-53:11:24.77 	&	&2.74	&	1.24	&				&	--	&	--	&				\tablenotemark{a}			\\
mm-10	&	16:01:44.62 &	-53:11:27.17	&	&8.88	&	0.67	&				&	11.74	&	1.42	&					\\
\hline
     \multicolumn{10}{c}{AGAL333} \\ \hline 
mm-1	&	16:20:56.83 	&	-50:44:01.98	&	&0.32	&	0.31	&				&	0.33	&	0.43	&							\\
mm-2	&	16:20:56.01 	&	-50:43:59.88 	&	&1.77	&	0.61	&				&	1.81	&	0.74	&							\\
mm-3	&	16:20:55.31 	&	-50:44:01.69 	&	&0.46	&	0.53	&				&	0.31	&	0.50	&							\\
mm-4	&	16:20:54.99 	&	-50:44:07.39 	&	&3.12	&	1.22	&				&	2.44	&	1.23	&							\\
mm-5	&	16:20:54.43 	&	-50:44:03.49 	&	&4.25	&	0.91	&				&	3.76	&	0.97	&							\\
mm-6	&	16:20:53.83 	&	-50:44:17.29 	&	&1.38	&	0.50	&				&	1.41	&	0.60	&							\\
mm-7	&	16:20:53.48 	&	-50:44:09.19	&	&0.58	&	0.78	&				&	0.60	&	0.77	&							\\
mm-8	&	16:20:52.28 	&	-50:44:18.78	&	&0.57	&	0.39	&				&	0.59	&	0.50	&							\\
mm-9	&	16:20:52.02 	&	-50:44:20.28 	&	&0.42	&	0.34	&				&	0.43	&	0.45	&							\\
mm-10	&	16:20:51.52 	&	-50:43:45.18	&	&0.68	&	0.87	&				&	0.22	&	0.45	&					 	\\
mm-11\tablenotemark{b}	&	16:20:51.71	&	-50:43:46.62	&	&	--	&	--	&				& 0.25	&	0.43	&					\tablenotemark{b}	\\
 \hline
\end{tabular}
\end{center}
\tablenotetext{a}{ Core not detected by Astrodendro.}
\tablenotetext{b}{ Core not detected by Clumpfind.}
\label{table:T3}
\end{table*}

In order to be quantitative in the identification of structures in the images we used two commonly 
employed methods: {\it Astrodendro\footnote{http://www.dendrograms.org/}} and {\it Clumpfind} (\citealt{clumpfind}).
We note that, in general, the number of extracted features and their parameters depend on the applied method (eg., \citealt{Pineda2009}). 
To identify structures in the {\it Astrodendro} algorithm, based on a dendrogram analysis (\citealt{Rosolowsky2008}), requires three inputs parameters: the minimum flux to be considered (F$_{min}$), the separation between neighboring peaks ($\delta$) and the minimum number of pixels (A$_{min}$) an structure should have. For a robust extraction of structures we adopted F$_{min}$=3$\sigma$, $\delta$=1$\sigma$ and A$_{min}$ = 1 beam. The key characteristic of this algorithm is its ability to track hierarchical structures over a range of scales.
The {\it Clumpfind} algorithm (\citealt{clumpfind}) is based on contouring the data array at different levels. The three input parameters are: the minimum flux level to be considered (T$_{low}$), the contour step ($\Delta$T) and the minimum number of pixels (S$_{min}$) required to be defined as a unique substructure. For core extraction we adopted T$_{low}$=3$\sigma$, $\Delta$T=2$\sigma$ and S$_{min}$ = 1 beam.

Towards AGAL329, Astrodendro identified 9 cores while Clumpfind recovered 10.  Towards AGAL333, Astrodendro identified 11 cores while Clumpfind identified 10 cores. The list of cores and their observed parameters are presented in Table \ref{table:T3}. Cols. 2 and 3  give the peak position, cols. 4 and 5 give, respectively,  the flux densities and deconvolved angular sizes (HWHM) determined from Clumpfind and cols. 6 and 7 those determined using Dendrogram. We find that the flux densities and angular sizes of the structures (cores) obtained from both methods are similar. Given the similarities, in the remaining of this paper we will use the parameters of the cores determined from the Clumpfind method (labeled in Figure \ref{fig:almaimages}).

\subsection{Molecular line emission}

\begin{figure*}
\centering
\includegraphics[trim={0.cm 0cm 0cm 0cm}, clip, width=0.70\textwidth]{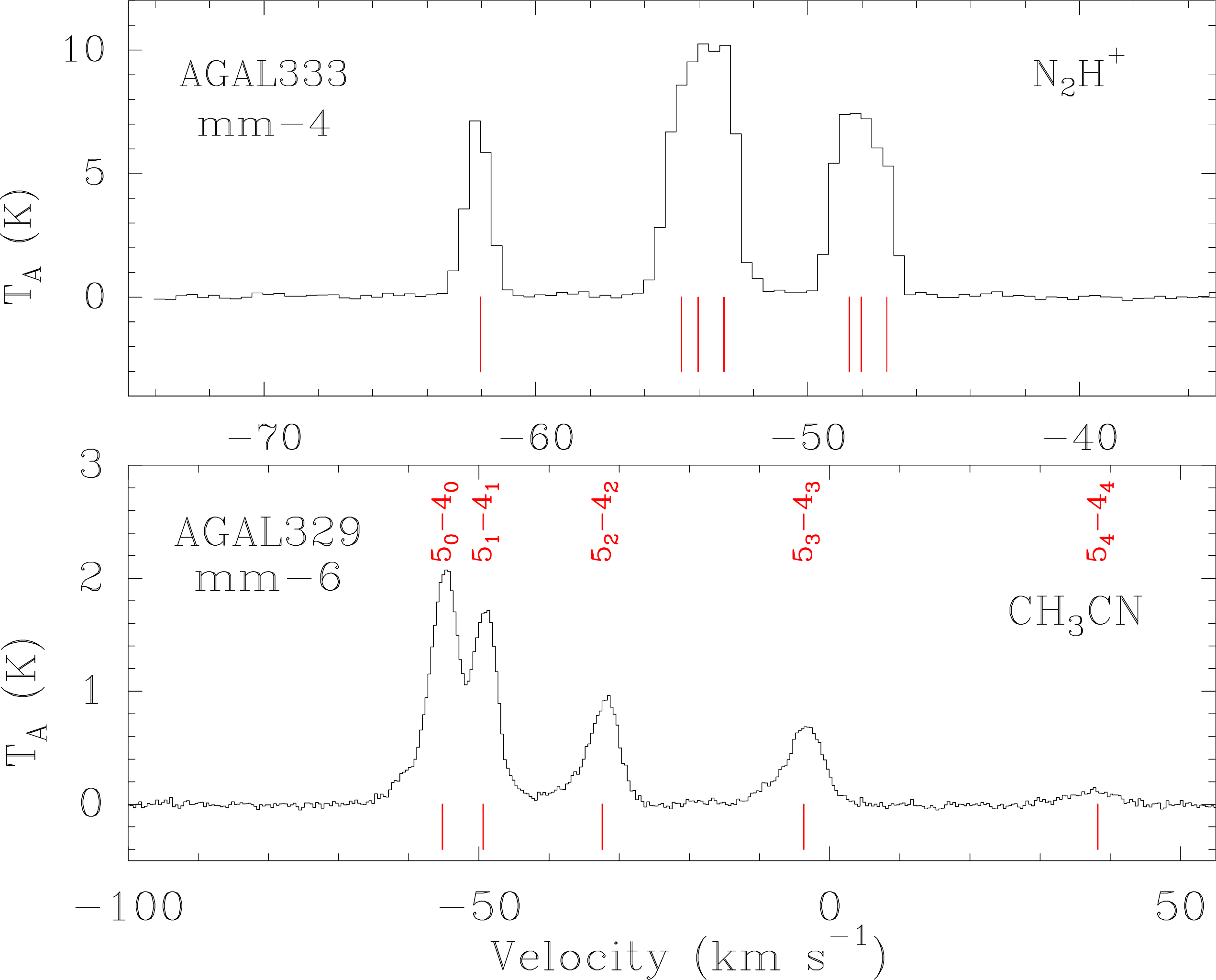}
\caption{Top panel: N$_2$H$^+$(J=1$\rightarrow$0) spectrum from core mm-4 in AGAL333. The red vertical lines at the bottom mark the velocities of the seven hyperfine components.  Bottom panel: CH$_3$CN($J=5\rightarrow4$) spectrum from core mm-6 in AGAL329. The red vertical lines at the bottom mark the velocities 
of the five K components, labeled at the top.  \label{eg_spec} }
\end{figure*}

Molecular line emission was detected in all four observed species towards both MDCs.  We note that the spectrum of the J=1$\rightarrow$0 transition of N$_{2}$H$^+$ consists of 7 hyperfine (HF) components (\citealt{Caselli1995}),
however, due to the overlap of closely spaced HF components, only 3 distinct lines are observed. This is illustrated in  Figure \ref{eg_spec} which shows the N$_{2}$H$^+$ spectrum observed toward core mm-4 in AGAL333. The lower velocity component of these three lines, centered at the frequency of 93176.265 MHz,  corresponds to a single HF component whereas the other two lines are blends of HF components. Also shown in Figure \ref{eg_spec} is the spectrum of the rotational J=5$\rightarrow$4 transition of CH$_{3}$CN observed toward core mm-6 in AGAL329. This rotational transition consists of 5 K components (marked in red), with K being the projection of the total angular momentum of the molecule about the principal rotation axis of the molecule. Their line frequencies, upper state energy levels and line strengths are given in Table \ref{table:ch3cnlinefreqs}. 

\begin{table*}[h!]
\begin{center}
\caption{CH$_{3}$CN  $J=5\rightarrow4$ rotational lines. \label{table:ch3cnlinefreqs}}
\begin{tabular}{cccccc}\hline
\hline 
	\multicolumn{1}{l}{Transition} &
	\multicolumn{1}{c}{ Frequency} &
	\multicolumn{1}{c}{ Velocity shift \tablenotemark{a}} &
	\multicolumn{1}{c}{ Eu/$k$ }  &
	\multicolumn{1}{c}{ Strength } &
	\multicolumn{1}{c}{S(I,K)} \\ 
 						& (MHz)			& (km s$^{-1}$)	&		(K)		&	$(J^2-K^2)/J$	& $g_k$$g_I$	\\ \hline 
$5_0\rightarrow4_0$		&	91987.09	&	....... 	& 	13.2		&	5.0	&	1/2	\\
$5_1\rightarrow4_1$		&	91985.31	&	5.79	&	20.4		&	4.8  &		1/2	\\
$5_2\rightarrow4_2$		&	91979.99	&	23.14	&	41.8		&	4.2	&		1/2	\\
$5_3\rightarrow4_3$		&	91971.13	&	52.04	&	77.5		&	3.2	&	1	\\
$5_4\rightarrow4_4$		&	91958.73	&	92.50	&	127.6	&	1.8	 & 1/2		\\ \hline
\end{tabular}
\end{center}
\tablenotetext{a}{ Shift with respect to the $5_0\rightarrow4_0$ line}
\end{table*}

To describe the emission at each position within the MDCs we performed moment analysis of the data \citep{Sault1995}, computing the zeroth  (integrated intensity), first (intensity weighted velocity field) and second  (intensity weighted velocity dispersion) moments. This approach allows an easy comparison of the characteristics of the emission in the different molecular transitions. The moments of the N$_{2}$H$^+$  emission were computed using the lower velocity component  of the 
three observed lines because it corresponds to a single HF component.  To make moment maps of the CH$_{3}$CN J=5$\rightarrow$4  emission towards  AGAL329 we used the emission observed in the $K = 2$ component  which is the stronger unblended component. 
Emission in this line was not detected towards AGAL333 and therefore we used the emission in the $5_0\rightarrow4_0$ component for the moment analysis. 

\subsubsection{Morphology}

\begin{figure*}[htbp!]
\centering
\includegraphics[trim={1cm 2cm 5cm 3.cm}, clip, width=0.98\textwidth]{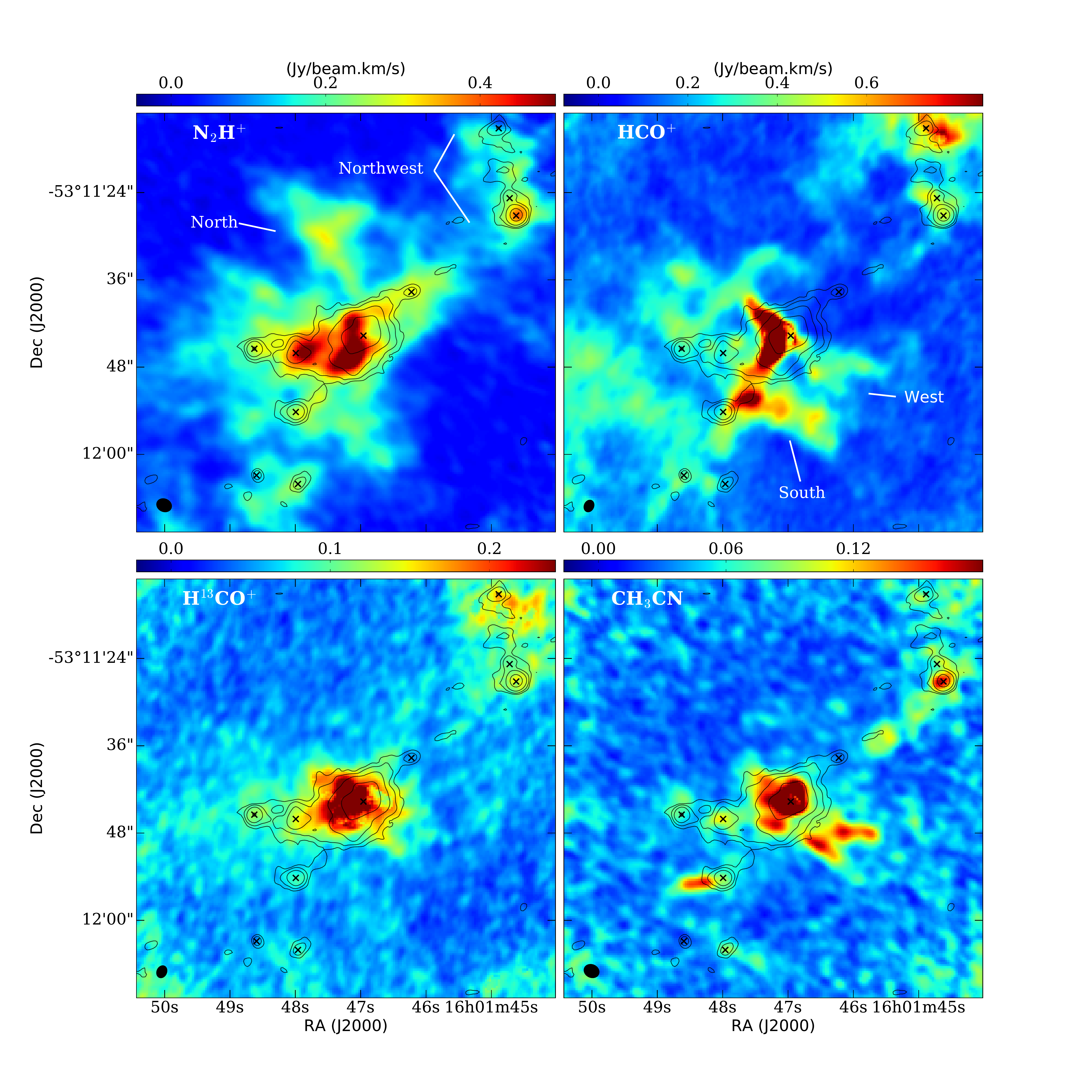}
\caption{Images of the velocity integrated line emission towards AGAL329. Superimposed are contours of the continuum 
emission.  Black crosses mark the peak position of the continuum cores. The black ellipse shown at the bottom left corner indicates the beam size. Top left: N$_{2}$H$^+$; top right: HCO$^{+}$, bottom left: H$^{13}$CO$^{+}$, bottom right: CH$_{3}$CN.
Labeled in the  different panels are conspicuous features discussed in the text. }
\label{agal329mom} 
\end{figure*}

\begin{figure*}
\centering
\includegraphics[trim={0.cm 0cm 0cm 0.cm}, clip,width=0.8\textwidth ]{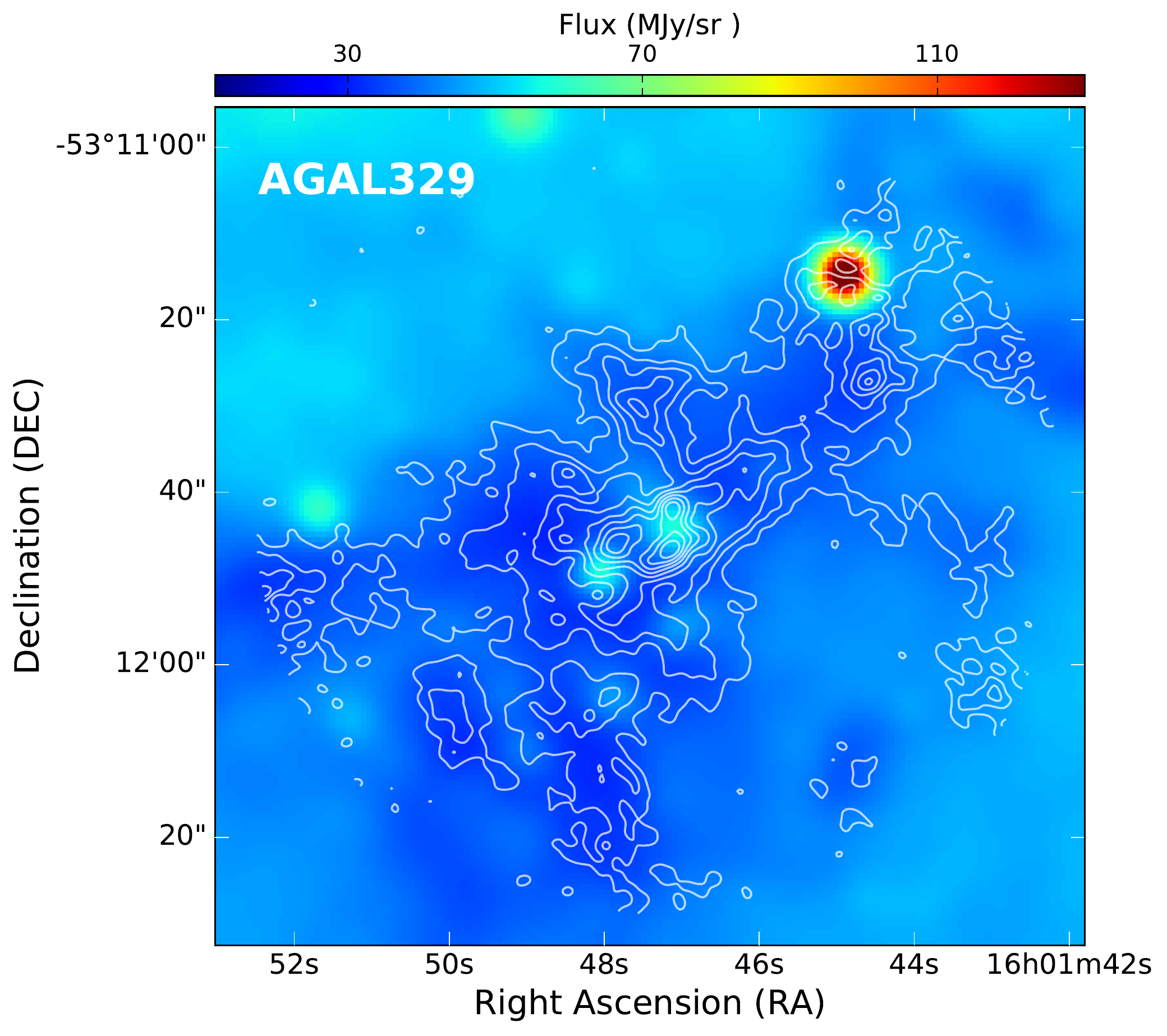}\caption{Spitzer 8$\micron$ image towards AGAL329 overlaid with contours of the velocity integrated N$_{2}$H$^{+}$ emission observed with ALMA. Contour levels are drawn from 10\% to 90\% of the peak emission of 0.70 Jy beam$^{-1}$ km s$^{-1}$, with a step of 10\%.  \label{n2hp_8mic}}
\end{figure*}

{\bf AGAL329:}
Figure \ref{agal329mom}  show images of the velocity integrated line emission (moment 0) in all four observed species towards AGAL329. 
The peak position of the continumm cores are marked with crosses. The velocity range of integration is from -84.0 to -32.0 km s$^{-1}$ for HCO$^{+}$, -64.0 to -40.0 km s$^{-1}$ for H$^{13}$CO$^{+}$, from -64.0  to -54.0 km s$^{-1}$ for N$_{2}$H$^+$ (corresponding to the lower velocity component of the hyperfine structure) and from -41.0 to -23.0 km s$^{-1}$ for the CH$_{3}$CN corresponding to the 
 J$_K = 5_2\rightarrow4_2$ component. The morphology of the line emission is noticeably different in the four transitions, most likely due to differences in optical depths,  excitation conditions and chemistry. 

The emission in the N$_{2}$H$^+$ line (upper left panel) is the brightest and most extended one of the four observed species. It exhibits a bright central region, with a radius of $\sim 6\arcsec$, surrounded by weaker emission from a region of $\sim 20\arcsec$ in radius, and an extended region emission seen toward the northwest of the image (labeled Northwest) which is associated with the GLIMPSE source G329.1845-00.3045 (\citealt{Robitaille2008}).  The bright central region exhibits a clumpy ring-like structure with three distinct condensations. The two westernmost condensations are associated with the mm-6 core, but their peak positions do not agree with the peak position of the continuum source, and the easternmost condensation is associated with the mm-4 core. All continuum cores are associated with N$_{2}$H$^{+}$ emission.  A conspicuous feature of the N$_{2}$H$^+$ image, is a region $\sim15\arcsec$  north of the central cores (labeled North), with a size of $\sim11\arcsec$ in diameter, which is not present in the other images and does not harbor continuum sources. Figure \ref{n2hp_8mic} presents an Spitzer image of the 8$\mu$m emission towards AGAL329, which clearly shows that this MDC is associated with an infrared dark cloud, superimposed with contours of the N$_{2}$H$^{+}$  emission. 
The morphology of the later closely follows the 8$\mu$m dark features, indicating that N$_{2}$H$^{+}$ is tracing gas with high column densities.  Interestingly the North N$_{2}$H$^+$ region is well correlated with an 8$\mu$m dark feature. This, together with  lack of  emission in the HCO$^{+}$, H$^{13}$CO$^{+}$ and CH$_{3}$CN lines suggests that this region is composed of dense and cold gas, which has undergone high levels of depletion.

The morphology of the HCO$^{+}$ emission exhibits noticeable differences with respect to that of the N$_{2}$H$^{+}$ emission. 
Towards the central N$_{2}$H$^{+}$ region, the HCO$^{+}$ emission shows a banana-like morphology which is roughly coincident with the 2 westernmost N$_{2}$H$^{+}$ condensations, but no HCO$^{+}$ emission is seen from the eastern
N$_{2}$H$^{+}$ condensation. The peak position of the mm-6 core is located at the western edge of the banana. Towards the extended Northwest region, the brighter HCO$^{+}$ emission is seen at its northern end (core mm-8) while the brighter N$_{2}$H$^{+}$ emission is seen at its southern end (core mm-10). In addition, the HCO$^{+}$ image shows two conspicuous features: a bright clumpy structure, located $\sim$18$\arcsec$ south of the central region, elongated in the NE-SW direction (labeled South),
barely seen in N$_{2}$H$^{+}$, and a weak V shaped feature located $\sim10\arcsec$  west from the central region (labeled West), not seen in the N$_{2}$H$^{+}$ image.
 
The H$^{13}$CO$^{+}$ emission exhibits a bright central component, with a size of 5.5$\arcsec$, which encompasses mm-6, 
diffuse emission seen towards the east, similar in extent to that seen in  N$_{2}$H$^{+}$,  and emission from the Northwest region.

\begin{figure*}[htp!]
\centering
\includegraphics[trim={0.5cm 2cm 5cm 3.cm}, clip, width=0.98\textwidth]{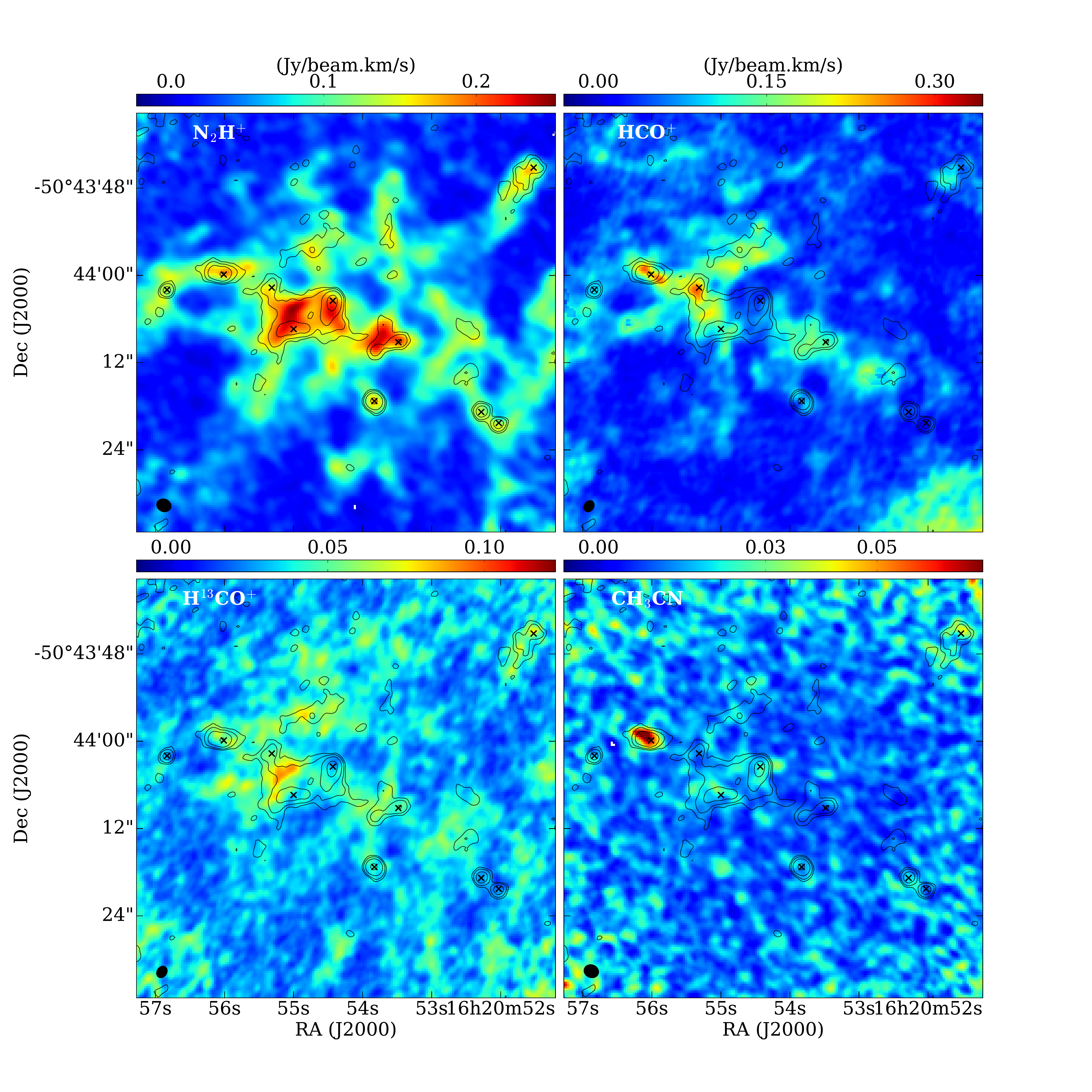}
\caption{Images of the velocity integrated line emission towards AGAL333. Superimposed are contours of the continuum 
emission.  Black crosses mark the peak position of the continuum cores. The black ellipse shown at the bottom left corner indicates the beam size.
Top left: N$_{2}$H$^+$; top right: HCO$^{+}$, bottom left: H$^{13}$CO$^{+}$, bottom right: CH$_{3}$CN.}
\label{agal333mom} 
\end{figure*}

The most prominent features in the CH$_{3}$CN image are a bright central region, with a size of 6$\arcsec$, whose peak position coincides with the peak position of core mm-6, a  bright V shaped region coincident with the West region seen in HCO$^{+}$ and an elongated, clumpy structure of weak emission running from northwest to southeast  which is closely associated with dark lanes seen in the 8 $\mu$m Spitzer images. We note here that the moment zero map of the CH$_{3}$CN emission was made using the emission in the K = 2 component in AGAL329 in order to avoid blending effects. The emission in the lower K components is much brighter and extended than in the higher K components. 

Figure \ref{agal329mom} also shows clear differences in the strength of emission from the cores in the different lines, likely caused 
by differences in optical depths, excitation conditions and/or  chemistry. The differences are illustrated by considering the three cores located in the northwest region of the clump: core mm-8 shows bright emisssion in HCO$^{+}$ and 
H$^{13}$CO$^{+}$,  core mm-10 is brighter in N$_{2}$H$^+$ and CH$_{3}$CN, while core mm-9 is brighter in HCO$^{+}$.

{\bf AGAL333:}
Figure \ref{agal333mom} presents images of the velocity integrated emission (moment 0) in all four observed species toward AGAL333,  showing that the morphology of the line emission is different in the four molecules. 
The peak position of the continumm cores are marked with crosses.
The velocity range of integration is -60.0 to -48.0 km s$^{-1}$ for HCO$^{+}$ and H$^{13}$CO$^{+}$, -67.0 to -60.0 km s$^{-1}$ for N$_{2}$H$^+$ and from -64.0 to -57.0 km s$^{-1}$ for the CH$_{3}$CN corresponding to the J$_K = 5_0\rightarrow4_0$ component. 
The emission in N$_{2}$H$^{+}$ line is the brightest and most extended one, delineating a complex network of filamentary structures across the whole region.  The main structure is a clumpy filament running from northeast to southwest, P.A. of 60 degrees. 
All of the continuum cores are associated with N$_{2}$H$^{+}$ emission and most of them lie within the main filament. There is a high degree of correlation between the N$_{2}$H$^{+}$ and the continuum emissions. 

The HCO$^{+}$ emission (upper right panel) clearly delineates the main filament running from northeast to southwest.
The brighter peaks of the HCO$^{+}$ emission are associated with cores mm-2 and mm-3.
The morphology of the H$^{13}$CO$^{+}$ emission shows some similarities to that of N$_{2}$H$^{+}$,  exhibiting a network of filamentary structures. However, the peak position of the brighter H$^{13}$CO$^{+}$ structures do not coincide with those of the continuum cores. In fact there is an anticorrelation between the H$^{13}$CO$^{+}$ and continuum emissions. The brighter feature in both the H$^{13}$CO$^{+}$ and N$_{2}$H$^{+}$ images corresponds to a region in between cores mm-3 and mm-4. Finally, emission in the CH$_{3}$CN line (lower right panel) was clearly detected only towards core mm-2 core and weakly detected towards cores mm-1, 3, 4 and 5. 

\subsubsection{Velocity field}

\begin{figure*}[htbp!]
\centering
\includegraphics[trim={0.cm 0cm 0cm 0cm}, clip, width=0.99\textwidth]{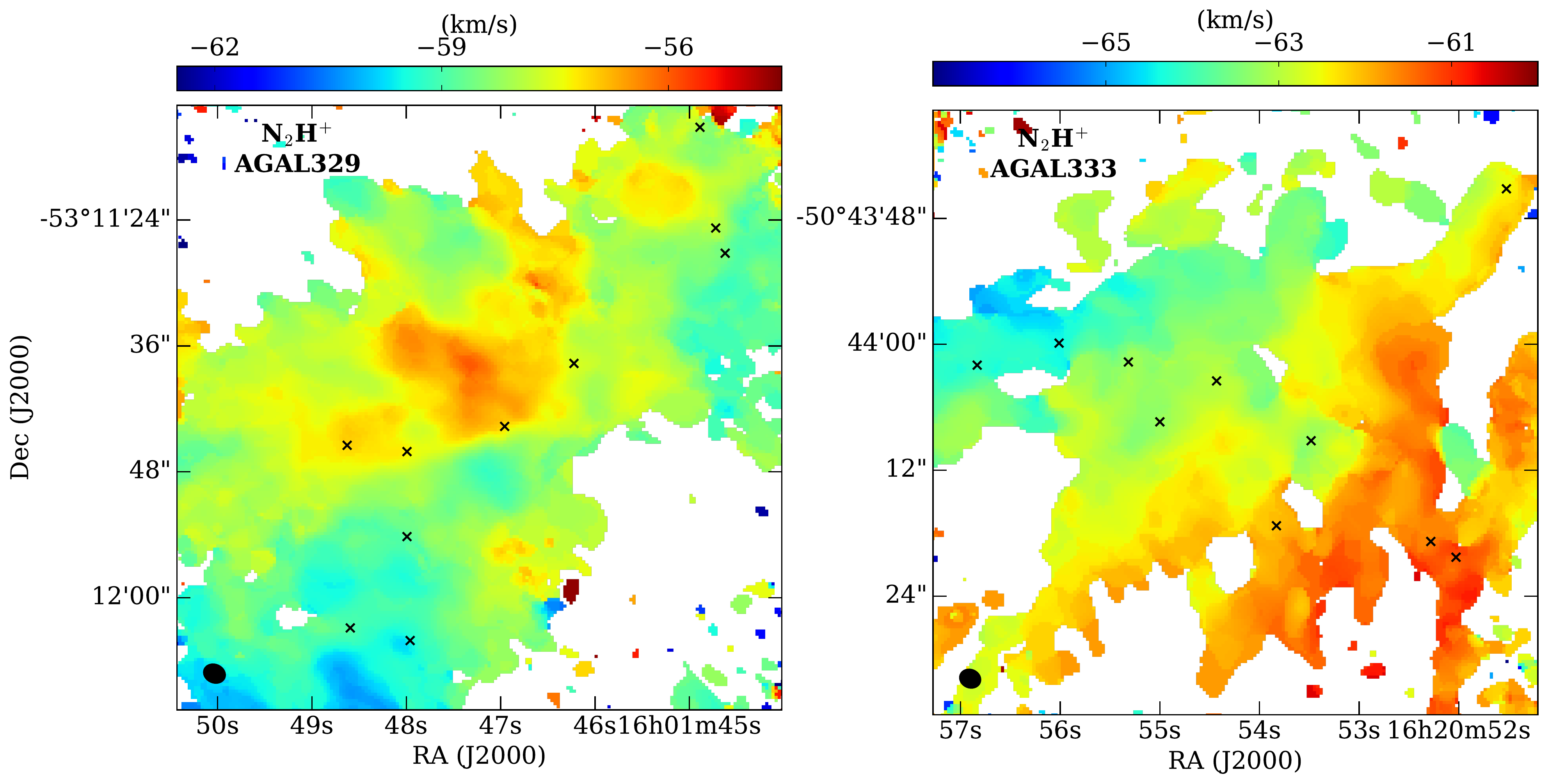}
\caption{Moment one images of the N$_{2}$H$^{+}$ emission from AGAL329 (left) and AGAL333 (right).  Crosses mark the peak position of the continuum cores.}
\label{agal333+329mom1} 
\end{figure*}

In order to investigate the velocity field across the MDCs we consider the emission in the N$_{2}$H$^{+}$  line which is bright and optically thin and therefore less affected by self-absorption effects. Figure \ref{agal333+329mom1} shows images of the velocity field (moment 1) of the N$_{2}$H$^{+}$  emission towards AGAL329 and AGAL333. 

Figure \ref{agal333+329mom1} (right panel) shows that there is a significant velocity gradient from northeast to southwest
across AGAL333. 
There is a velocity shift of 4.2 km s$^{-1}$ over a region of about 50$\arcsec$, which at the distance of 3.72 kpc corresponds to a velocity gradient of 4.7 km s$^{-1}$  pc$^{-1}$.
Assuming that the velocity gradient is due to gravitationally bound rotation of an structure with radius R and mass M, then 
M = (dV/dR)$^{2} R^3/G$ (c.f., \citealt{Armstrong1985}). The observed velocity gradient implies a mass 
within a radius of 0.45 pc of 460 M$_\odot$. This mass is within a factor of two from the mass derived from the dust observations, giving support to the bound rotation hypothesis.

\begin{figure*}[!hbt]
\centering
\includegraphics[trim={5.cm 1.65cm 9.1cm 3.65cm}, clip, width=0.99\textwidth]{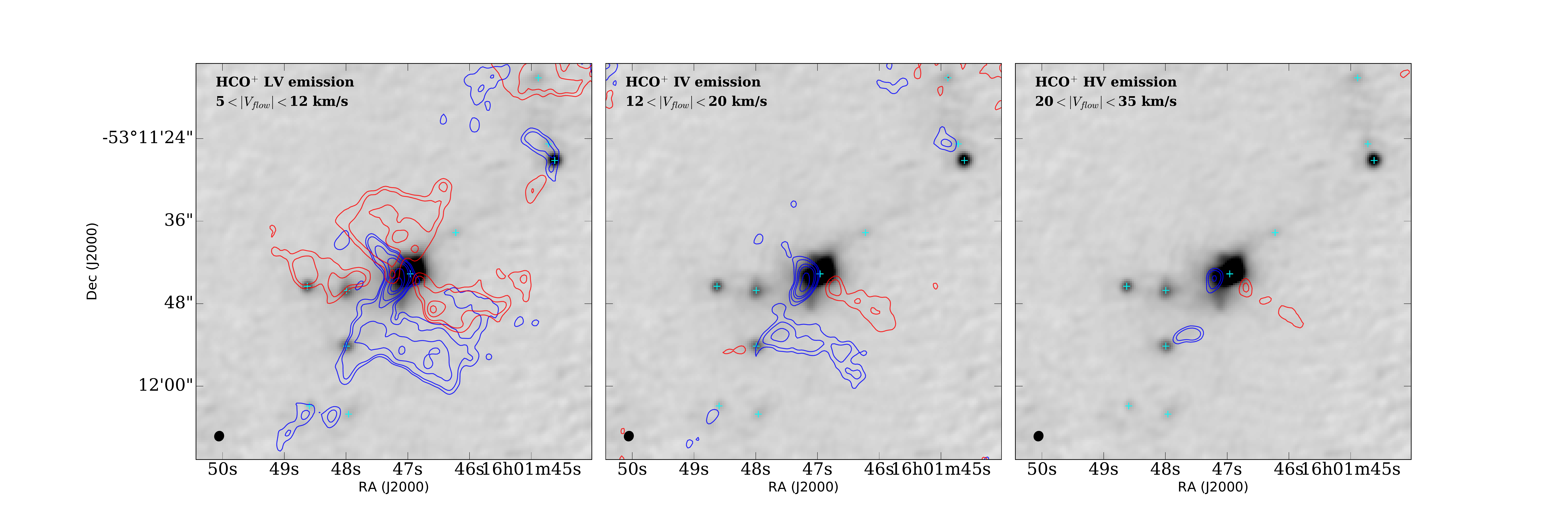}
\caption{Contour maps of low velocity (LV), intermediate velocity (IV) and high velocity (HV) HCO$^{+}$ emission towards AGAL329 overlaid in the 3 mm ALMA continuum map (gray scale). The blue and red color contours mark blue-shifted and red-shifted emission, respectively. The flow velocity range is shown in the top left corner of each map. }
\label{outflow} 
\end{figure*}

The velocity field towards the protostellar clump (Figure \ref{agal333+329mom1}, left panel) appears complex, with no organized motions nor clear velocity gradients seen across the clump. The redder velocities seen towards the north and the bluer velocities seen towards the south are probably caused by the presence of outflows, as discussed next. In several positions across this clump the profiles of the HCO$^{+}$ line emission exhibit the presence of  wing emission. To investigate the spatial distribution of the wing emission, we made contour maps of the velocity integrated emission in three ranges of  radial flow velocities. The radial flow velocity, v$_{flow}$, is defined as $v_{LSR}-v_{0}$, where $v_{0}$ is the systemic velocity of the clump, assumed to be -50.5 km s$^{-1}$.  Figure \ref{outflow} shows maps of the wing emission, overlaid on the ALMA dust continuum image,  in three ranges of flow velocities: 20 $< |v_{flow}| <$  35  km s$^{-1}$, referred as the high velocity (HV) wing,  
12 $< |v_{flow}| <$  20  km s$^{-1}$, referred as the intermediate velocity (IV) wing, and
 5 $< |v_{flow}| < $ 12  km s$^{-1}$, referred as the low velocity (LV) wing.
The morphology of the HCO$^{+}$ wing emission is complex. Clearly distinguished in the LV map is an extended bipolar-like structure with a wide opening angle (half opening angle of $\sim$43$^{\circ}$), consisting of a lobe of redshifted emission seen towards the north  and a lobe of blueshifted emission seen towards the south, located on opposite directions from core mm-6. The position angle of the symmetry axis of the outflow is  P.A. $\sim$5 degrees. The linear extensions of the redshifted and blueshifted lobes along the symmetry axis are $\sim$0.23 pc ($\sim$14$\arcsec$) and $\sim$0.27 pc ($\sim$16$\arcsec$), respectively. Also distinguished in the LV map is a 
second, more collimated, bipolar-like structure, with a position angle of 45 degrees, consisting of a redshifted lobe 
extending toward the southwest and a blueshifted lobe extending toward the northeast from core mm-6.
The blueshifted lobe extends $\sim$0.12 pc ($\sim$7$\arcsec$) northeast while the redshifted lobe extends $\sim$0.25 pc ($\sim$15$\arcsec$) southwest. In addition, seen in the LV map is a weak blueshifted feature extending towards the northeast and a weak redshifted feature extending towards the southeast from core mm-10. These features may correspond to streams of gas infalling towards core mm-10. 

In the IV map, emission from the wide angle bipolar structure is only seen from the blueshifted lobe. In this velocity range emission 
from the more collimated bipolar structure is clearly seen at redshifted velocities.
In the HV map, the blueshifted emission associated with the mm-core 6 is compact ($\sim$3$\arcsec$) while the redshifted emission extending west shows three separate `knot' like features, at distances of $\sim$3$\arcsec$, $\sim$6.5$\arcsec$ and $\sim$10$\arcsec$ from peak position of mm-6.  It is notable that this emission region is also detected in CH$_{3}$CN emission (see Figure \ref{agal329mom}, labeled West). Also seen in HV map is a narrow blueshifted emission, of size $\sim$4$\arcsec$, associated with mm-3.  

\subsection{Line emission from cores}

\begin{figure*}[htb]
\centering
\includegraphics[trim={0.cm 0cm 0cm 0cm}, clip, width=0.95\textwidth]{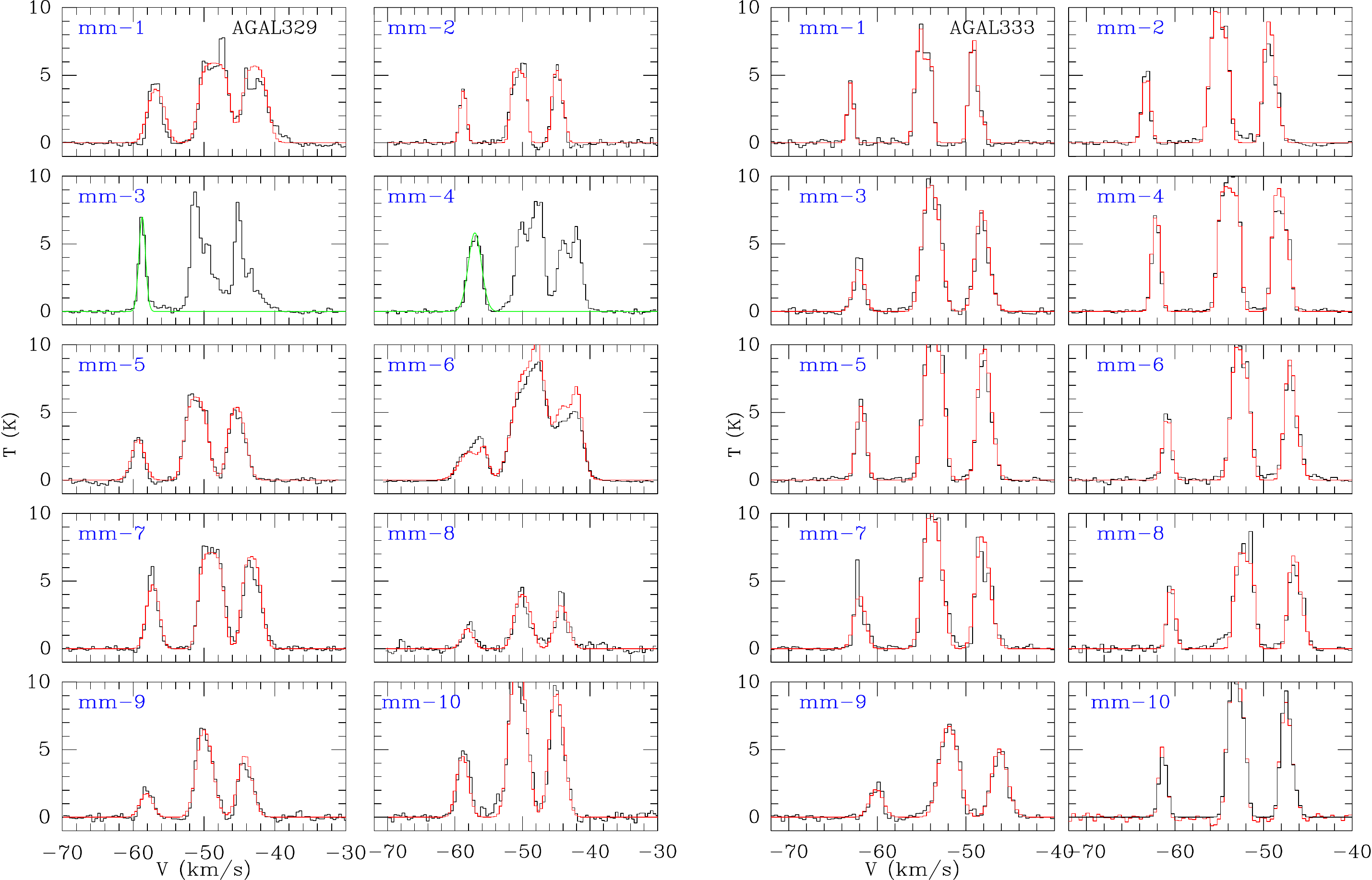}
\caption{Average spectra of the N$_{2}$H$^{+}$ emission from cores. Left panel:  AGAL329 cores. Right panel: AGAL333 cores. The red line indicates the result of a simultaneous fit to the whole hyperfine structure and the green line indicates the result of a Gaussian fit to the lower velocity component.}
\label{spec_n2hp} 
\end{figure*}

\begin{figure*}[htb]
\centering
\includegraphics[trim={0cm 0cm 0cm 0cm}, clip, width=0.95\textwidth]{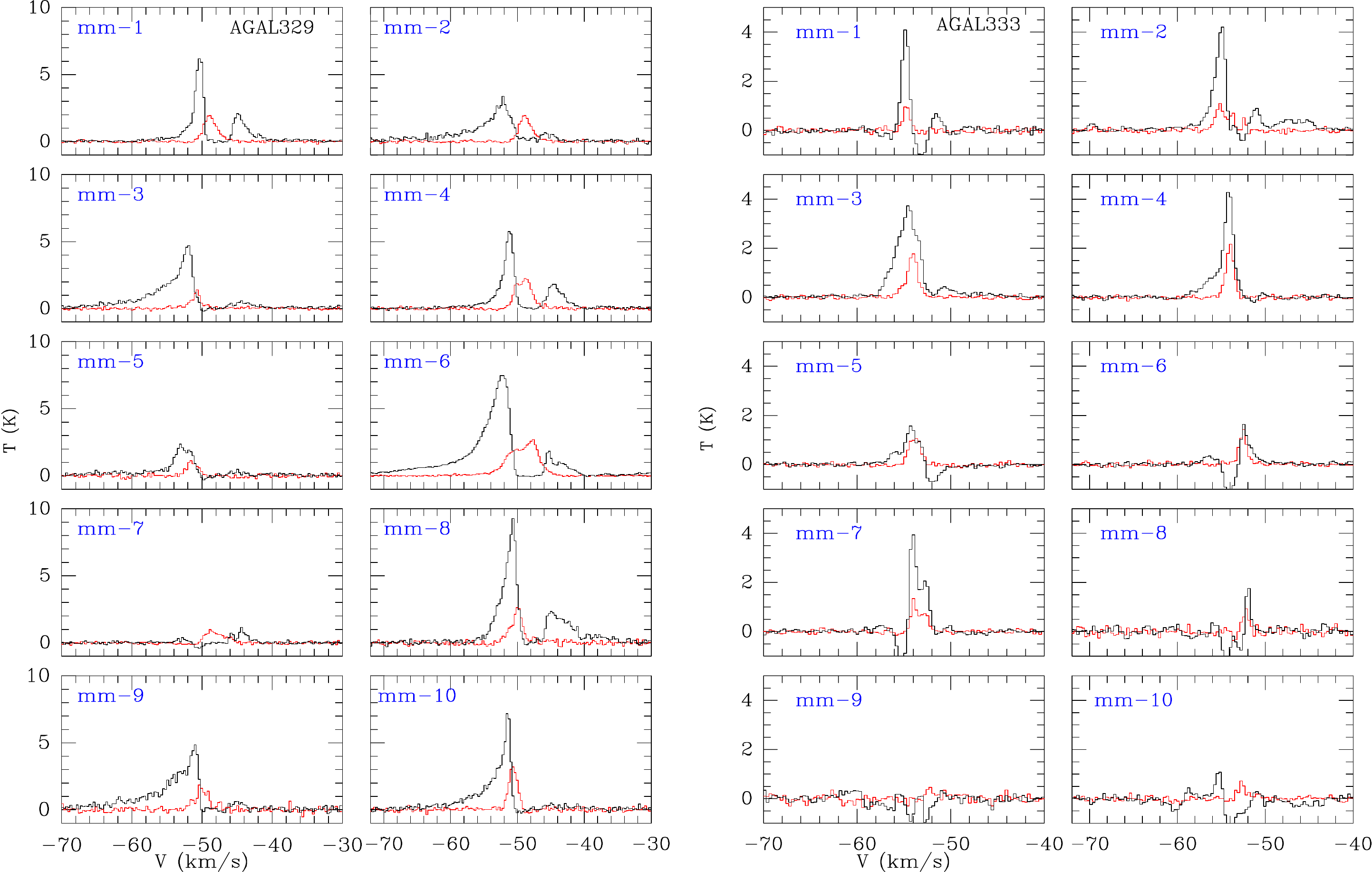}
\caption{Average spectra of the HCO$^{+}$ (black line) and H$^{13}$CO$^{+}$ (red line) emission from cores. Left: AGAL329 cores. Right: AGAL333 cores. \label{spec_hcop}}
\end{figure*}

\begin{figure*}[htb]
\centering
\includegraphics[trim={0cm 0cm 0cm 0cm}, clip, width=0.95\textwidth]{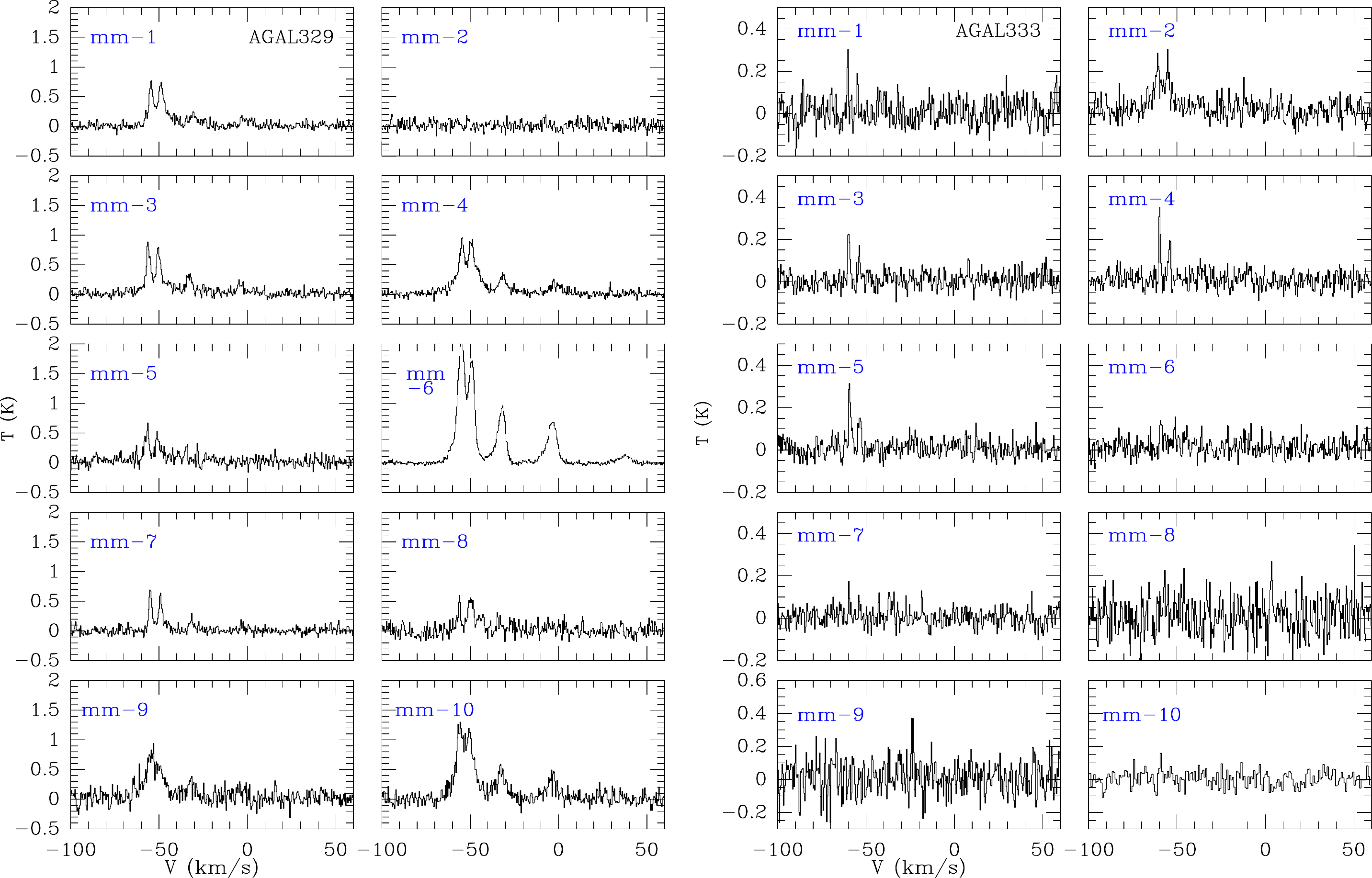}
\caption{Average spectra of the CH$_{3}$CN  emission from cores. Left: AGAL329 cores. Right: AGAL333 cores.}
\label{spec_ch3cn} 
\end{figure*}

In this section we present the characteristics of the spectra of the molecular line emision from the continuum cores. 
The spectra correspond to the average spectra of the spatially integrated emission over the solid angle subtended by each core (hereafter refereed as the core spectra). 

Figure \ref{spec_n2hp}  shows the spectra of the N$_{2}$H$^{+}$ J=1$\rightarrow$0 line. Emission is clearly detected towards all cores in both MDCs. The red line shows the result of a simultaneous fit to the emission from the whole hyperfine structure for a single velocity component, made using the task {\it pyspeckit.models} in {\it PySpecKit} {\citep{Pyspekit}. This approach gave good results 
for all cores in AGAL333 and for most cores in AGAL329, except cores mm-3, mm-4 and mm-6 which exhibits either self-absorption profiles or strong wing emission. Thus, for cores mm-3 and mm-4 in AGAL329 we fitted a single gaussian profile to the lower velocity (single HF) component (green line), while for core mm-6, two velocity components were used to fit the spectra. 

Figure \ref{spec_hcop} shows, in the same panel,  the spectra of the HCO$^{+}$ (black line) and H$^{13}$CO$^{+}$ (red line) emission.  Emission in these lines was detected from all cores in both MDCs,  except toward core mm-9 in AGAL333 in which the H$^{13}$CO$^{+}$ line was not detected ($<$3$\sigma$). The HCO$^{+}$ profiles from several cores in AGAL329  display line asymmetries and self absorption features. Typically the HCO$^{+}$ profile shows two peaks with a strong blueshifted peak and weak redshifted peak relative to the velocity of the optically thin H$^{13}$CO$^{+}$ line. This is a characteristic signature of infalling gas, probably due to the global collapse of the clump (see \S\ref{collapsesignature}). 

Most cores within the AGAL333 prestellar clump  show absorption features and/or shoulders in the HCO$^{+}$ line. Cores mm-1, mm-2, mm-4, mm-5 and mm-10 show inverse P-Cygni profiles, suggesting that they are undergoing contraction motions. On the other hand cores mm-6, mm-7 and mm-8 show P-Cygni like profiles, usually taken as a signpost of outflowing or expanding gas motions.
It is possible that some of the cores formed in MDCs be transient objects and therefore be expanding. In fact the mm-7 core has a virial parameter of 2.5, further suggesting it is not bounded. 

Figure \ref{spec_ch3cn} presents the observed spectra in the $J_K = 5_K\rightarrow4_K$ CH$_{3}$CN K-ladder. Emission  was detected from all cores within AGAL329, except mm-2.  Emission was detected in all five K components towards one core (mm-6), in four K components towards five cores (mm-1, 3, 4, 7 and 10), in three K components towards two cores (mm-5 and mm-9), and in two K components towards one core (mm- 8). Towards  AGAL333, only weak CH$_{3}$CN emission was detected from cores mm-1, mm-2, mm-3, mm-4 and mm-5.

\subsubsection{Line parameters of optically thin transitions}

\begin{table*}
\begin{center}
\caption{Line parameters of the core emission in optically thin lines. \label{lwtable}} \smallskip
\footnotesize
\begin{tabular}{lhccccccc} 
	\hline \hline
	\multicolumn{2}{c}{Core} & 
	\multicolumn{3}{c}{ N$_{2}$H$^{+}$ $J$=1$\rightarrow 0$ }  & \multicolumn{1}{c}{} & 
		\multicolumn{3}{c}{ H$^{13}$CO$^{+}$ $J$=1$\rightarrow 0$ }  \\	\cline{2-5} \cline{7-9}	
    		& HFS 	&	V$_{lsr}$				&	$\Delta$V			&			$\tau_{tot}$		& 	&   V$_{lsr}$				&	$\Delta$V	 		&   T$_{A}$				\\ 
		& Comp.	&      km s$^{-1}$					&		km s$^{-1}$					&       			&       &	km s$^{-1}$					&	km s$^{-1}$			 	&       K	     		 \\ \hline 
\multicolumn{9}{c}{AGAL329} \\ \hline
mm-1	&	1	&	-48.71	$\pm$	0.04	&	2.35	$\pm$	0.09	&	10.0	$\pm$	1.4	&	&	-48.87	$\pm$	0.04	&	2.11	$\pm$	0.09	&	2.55	$\pm$	0.09	\\
mm-2	&	1	&	-50.72	$\pm$	0.02	&	1.01	$\pm$	0.05	&	10.1	$\pm$	1.3	&	&	-50.85	$\pm$	0.11	&	1.27	$\pm$	0.27	&	0.95	$\pm$	0.17	\\
mm-3$^a$	&	$^a$	&	-50.76	$\pm$	0.09	&	1.09	$\pm$	0.24	&	--				&	&	-50.92	$\pm$	0.03	&	1.00	$\pm$	0.07	&	2.25	$\pm$	0.13	\\
mm-4$^a$	&	$^a$	&	-49.05	$\pm$	0.10	&	2.29	$\pm$	0.02	&	--				&	&	-49.09	$\pm$	0.02	&	2.84	$\pm$	0.04	&	2.23	$\pm$	0.10	\\
mm-5	&	1	&	-51.24	$\pm$	0.02	&	2.00	$\pm$	0.05	&	5.2	$\pm$	0.5	&	&	-51.70	$\pm$	0.09	&	2.22	$\pm$	0.22	&	1.27	$\pm$	0.11	\\
mm-6	&	1	&	-49.82	$\pm$	0.20	&	3.48	$\pm$	0.35	&	1.1	$\pm$	0.4	&	&	-50.59	$\pm$	0.05	&	4.86	$\pm$	0.12	&	3.36	$\pm$	0.04	\\
		&	2	&	-47.61	$\pm$	0.05	&	1.36	$\pm$	0.21	&	0.5	$\pm$	0.1	&	&	-47.57	$\pm$	0.02	&	1.49	$\pm$	0.07	&	2.53	$\pm$	0.11	\\
mm-7	&	1	&	-49.15	$\pm$	0.02	&	1.81	$\pm$	0.05	&	9.6	$\pm$	0.9	&	&	-48.39	$\pm$	0.13	&	2.12	$\pm$	0.30	&	0.81	$\pm$	0.10	\\
mm-8	&	1	&	-50.13	$\pm$	0.05	&	1.86	$\pm$	0.14	&	2.7	$\pm$	1.0	&	&	-50.30	$\pm$	0.07	&	2.48	$\pm$	0.17	&	2.56	$\pm$	0.15	\\
mm-9	&	1	&	-49.71	$\pm$	0.03	&	2.72	$\pm$	0.07	&	2.2	$\pm$	0.4	&	&	-50.03	$\pm$	0.16	&	2.82	$\pm$	0.38	&	1.29	$\pm$	0.15	\\
mm-10	&	1	&	-50.83	$\pm$	0.02	&	1.97	$\pm$	0.07	&	3.6	$\pm$	0.5	&	&	-50.84	$\pm$	0.04	&	1.63	$\pm$	0.09	&	3.97	$\pm$	0.19	\\
\hline 
\multicolumn{9}{c}{AGAL333} \\ \hline
mm-1	&	1	&	-54.97	$\pm$	0.01	&	0.75	$\pm$	0.02	&	4.0	$\pm$	0.5	&	&	-54.94	$\pm$	0.06	&	0.81	$\pm$	0.14	&	1.14	$\pm$	0.17	\\
mm-2	&	1	&	-55.18	$\pm$	0.01	&	1.03	$\pm$	0.05	&	4.0	$\pm$	0.6	&	&	-55.16	$\pm$	0.09	&	1.86	$\pm$	0.21	&	1.23	$\pm$	0.12	\\
mm-3	&	1	&	-54.03	$\pm$	0.02	&	1.46	$\pm$	0.05	&	1.4	$\pm$	0.4	&	&	-54.13	$\pm$	0.05	&	1.51	$\pm$	0.11	&	1.61	$\pm$	0.10	\\
mm-4	&	1	&	-54.07	$\pm$	0.01	&	0.96	$\pm$	0.02	&	12.5	$\pm$	1.3	&	&	-54.24	$\pm$	0.02	&	0.91	$\pm$	0.06	&	2.13	$\pm$	0.12	\\
mm-5	&	1	&	-53.80	$\pm$	0.01	&	1.22	$\pm$	0.02	&	4.7	$\pm$	0.5	&	&	-53.80	$\pm$	0.05	&	1.11	$\pm$	0.11	&	1.31	$\pm$	0.11	\\
mm-6	&	1	&	-52.81	$\pm$	0.01	&	1.10	$\pm$	0.02	&	3.4	$\pm$	0.5	&	&	-52.78	$\pm$	0.04	&	0.93	$\pm$	0.09	&	1.70	$\pm$	0.14	\\
mm-7	&	1	&	-53.95	$\pm$	0.03	&	1.46	$\pm$	0.07	&	2.6	$\pm$	0.8	&	&	-54.01	$\pm$	0.03	&	1.02	$\pm$	0.07	&	1.70	$\pm$	0.10	\\
mm-8	&	1	&	-52.37	$\pm$	0.02	&	0.96	$\pm$	0.05	&	8.0	$\pm$	1.3	&	&	-52.32	$\pm$	0.10	&	0.86	$\pm$	0.24	&	0.91	$\pm$	0.22	\\
mm-9	&	1	&	-52.10	$\pm$	0.01	&	1.50	$\pm$	0.05	&	2.8	$\pm$	0.4	&	& --	& -- & --  \\ 
mm-10	&	1	&	-53.28	$\pm$	0.01	&	1.06	$\pm$	0.02	&	2.3	$\pm$	0.4	&	&	-53.07	$\pm$	0.04	&	0.95	$\pm$	0.10	&	1.91	$\pm$	0.17	\\
\hline
\end{tabular}

\tablenotetext{^a}{N$_{2}$H$^{+}$ line velocity and widths is obtained from the Gaussian fit to the lower velocity component. }
\end{center}
\end{table*}
\begin{figure*}[htb!]
\centering
\includegraphics[trim={0.cm 0cm 0cm 0cm}, clip, width=0.95\textwidth]{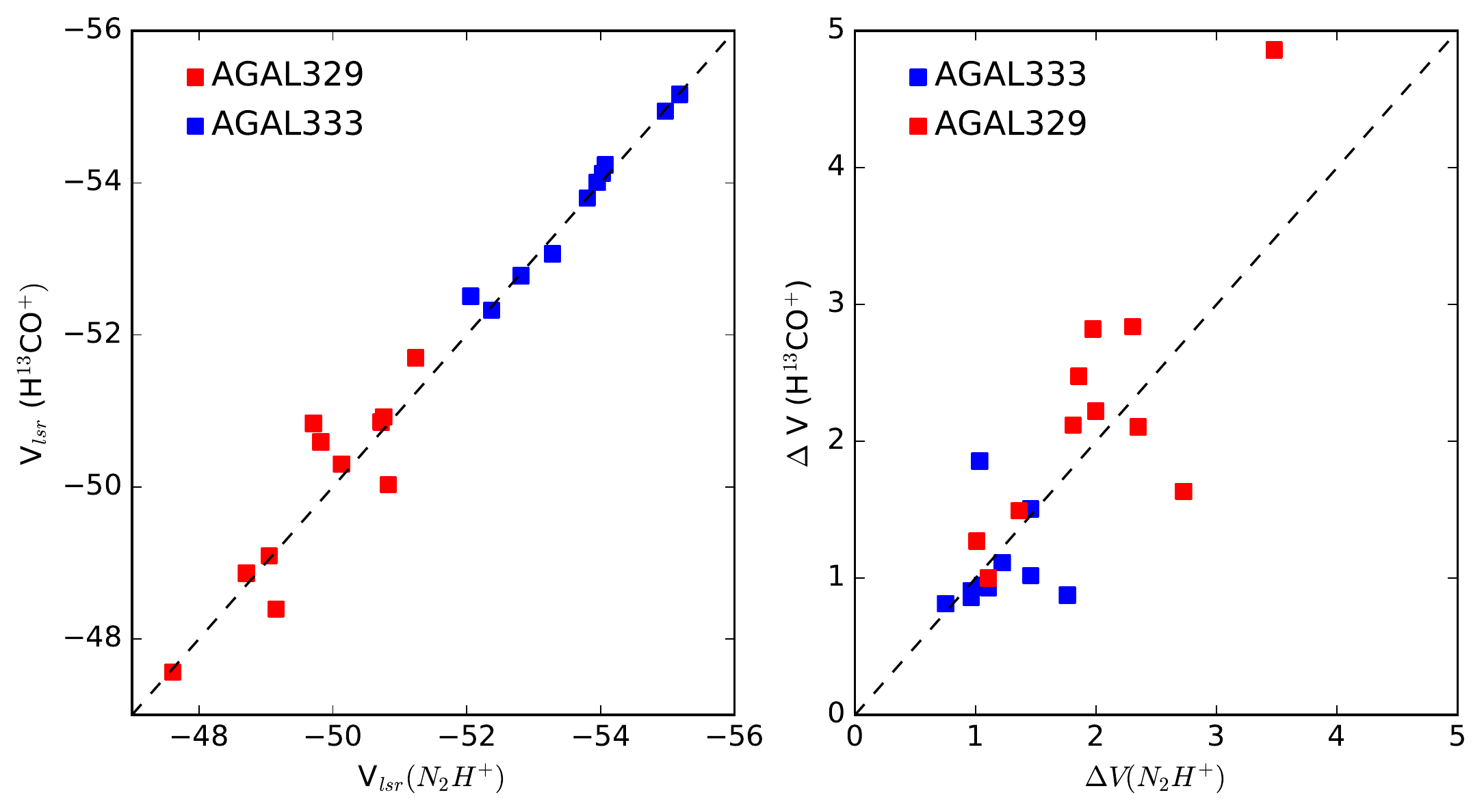}
\caption{Comparison between the line velocities (left panel) and line widths (right panel) derived from the N$_{2}$H$^{+}$ hyperfine fit and the H$^{13}$CO$^{+}$ gaussian fit. Blue  and red squares indicate values for cores in AGAL333 and in AGAL329, respectively. \label{lwdist}}
\end{figure*}

The determination of the kinematics (velocities), turbulence (line widths) and dynamical state of the cores (virial parameter) requires observations of the emission in optically thin molecular lines, since they are free of self-absorption features.
Table \ref{lwtable} lists the parameters of the core emission in the optically thin J=1$\rightarrow$0 lines of H$^{13}$CO$^{+}$ and N$_{2}$H$^{+}$. The former were determined from a Gaussian fit to the core spectra while the latter were derived in most cases from a simultaneous fit to all hyperfine lines. For core mm-6 in AGAL329 two velocity components were used to fit the N$_{2}$H$^{+}$ and H$^{13}$CO$^{+}$ profiles. In addition to the central velocity and linewidth, the simultaneous hyperfine fit provides the total optical depth in the N$_{2}$H$^{+}$ J=1$\rightarrow$0 transition. The derived total optical depths range from $\sim$1.4 to 12.5 for the cores in prestellar source and from $\sim$0.5 to 10.1 for the cores in protostellar source. Even though the total optical depths are $\ge$1, the individual hyperfine components are mostly optically thin. 
As shown in Figure \ref{lwdist}, the velocities and line widths of the cores derived from both lines are in good agreement. 

In general the line widths of the cores in the prestellar clump are smaller than those of the cores in the protostellar clump
(see  Figure  \ref{lwdist}). The average line width of the cores within AGAL333 and AGAL329 are 1.2 km s$^{-1}$ and 2.0 km s$^{-1}$, respectively. The explanation of the large linewidths in cores within the protostellar clump is not straightforward,  it may reflect either an increase in the level of turbulence due to the beginning of star formation activity or an increase in the gas velocities due to collapse motions.
In particular,  \cite{Vazquez2009} concluded that in a cloud undergoing global gravitational collapse, the velocity dispersion at all  scales are caused by infall motions rather than by turbulence.

\section{Analysis and discussion \label{sec:s5}}

In this section we discuss the physical parameters of the cores, their spatial distribution, virial state and the fraction of the total mass in cores relative to the parent clump mass, all of which are key properties to discern among the different models of fragmentation of MDCs. 

\subsection{Core parameters}

\begin{table*}[ht!]
\caption{Derived core parameters. \label{coreparam}}
\footnotesize
\begin{center}
\begin{tabular}{lccccccccc}
    \hline \hline
    \multicolumn{10}{c}{Derived parameters} \\  \hline
  Core   	&  	Temp.	&	Mass		& & 	\multicolumn{2}{c}{Radius}	 &	 \multicolumn{1}{c}{n(H$_{2}$)}	& $\sigma_{vir}$ \tablenotemark{$^\dagger$} & M$_{vir}$	 &	$\alpha$$_{vir}$  \\ \cline{5-6}
		&	(K)  		& 	(M$_{\odot}$)	& &	 (pc)	& (10$^{3}$AU)		&	 (10$^{7}$ cm$^{-3}$)  &  (km s$^{-1}$)&	(M$_{\odot}$)	&			\\ \hline 
\multicolumn{10}{c}{AGAL329}\\ \hline
mm-1	&	30	&	14.8	& &	0.011	&	2.3	&	3.9	&	1.05	&	14.0	&	0.95	\\
mm-2	&	28	&	1.7	& &	0.005	&	1.0	&	4.6	&	0.53	&	1.6	&	0.96	\\
mm-3	&	41	&	11.6	& &	0.012	&	2.5	&	2.3	&	0.59	&	4.9	&	0.42	\\
mm-4	&	34	&	27.2	& &	0.019	&	3.9	&	1.4	&	1.03	&	23.4	&	0.86	\\
mm-5	&	33	&	3.0	& &	0.008	&	1.7	&	2.0	&	0.91	&	7.7	&	2.60	\\
mm-6	&	68	&	118.8\tablenotemark{$^\ddagger$}&	&	0.030	&	6.2	&	1.5	&	1.55	&	84.2	&	0.71	\\
mm-7	&	31	&	3.1	& &	0.009	&	1.9	&	1.5	&	0.83	&	7.3	&	2.38	\\
mm-8	&	28	&	8.5	& &	0.013	&	2.7	&	1.3	&	0.85	&	10.9	&	1.28	\\
mm-9	&	28	&	9.3	& &	0.021	&	4.3	&	0.4	&	1.20	&	34.9	&	3.74	\\
mm-10	&	41	&	20.1	& &	0.011	&	2.3	&	5.3	&	0.91	&	10.7	&	0.53	\\\hline 
\multicolumn{10}{c}{AGAL333}\\ \hline
mm-1	&	22	&	1.6	& &	0.006	&	1.2	&	2.6	&	0.42	&	1.2	&	0.74	\\
mm-2	&	31	&	6.3	& &	0.011	&	2.3	&	1.6	&	0.54	&	3.8	&	0.60	\\
mm-3	&	22	&	2.4	& &	0.010	&	2.1	&	0.8	&	0.68	&	5.3	&	2.25	\\
mm-4	&	30	&	11.4	& &	0.022	&	4.5	&	0.4	&	0.51	&	6.8	&	0.59	\\
mm-5	&	24	&	19.8	& &	0.016	&	3.3	&	1.7	&	0.59	&	6.5	&	0.33	\\
mm-6	&	22	&	7.1	& &	0.009	&	1.9	&	3.4	&	0.54	&	3.0	&	0.43	\\
mm-7	&	22	&	3.0	& &	0.014	&	2.9	&	0.4	&	0.68	&	7.5	&	2.50	\\
mm-8	&	22	&	2.9	& &	0.007	&	1.4	&	3.0	&	0.49	&	1.9	&	0.66	\\
mm-9	&	22	&	2.2	& &	0.006	&	1.2	&	3.5	&	0.80	&	4.4	&	2.04	\\
mm-10	&	22	&	3.5	& &	0.016	&	3.3	&	0.3	&	0.52	&	5.1	&	1.46	\\\hline
\end{tabular}
\end{center}
\tablenotetext{^\dagger}{$\sigma_{vir} = \sqrt{({\sigma_{th}^{2} + \sigma_{tur}^{2}})}$,  where $\sigma_{th}$ is the thermal width and $\sigma_{tur}$ is the turbulent width.}
\tablenotetext{^\ddagger}{Computed from the observed flux density at 3mm corrected by the contribution from the free-free emission.}

\end{table*}
The physical parameters of the cores are listed in Table \ref{coreparam}
 and their derivation is discussed in what follows. 

\subsubsection{Sizes}
The radius of the cores were computed from the geometric mean of the semi-major and minor axis determined from Clumpfind and the distances given in Table \ref{sampleparams}. For cores in AGAL329 the radii range from 0.005  to 0.030 pc (1000 to 6200 AU) with an average value of 0.014 pc (2900 AU), while for cores in AGAL333 the radii range from 0.006 to 0.022 pc (1200 to 4500 AU) with an average value of 0.012 pc (2500 AU).  
 
\subsubsection{Temperatures \label{ch3cn_temp}}
The detection of emission in at least two J=5$_K\rightarrow4_K$ transitions of CH$_{3}$CN  allows to determine the rotational temperature of methyl cyanide, which is known to provide a good estimate of the kinetic temperature of the gas (eg., \citealt{Guesten1985,rejiman2004, hernandez-hernandez2014}). We use the standard rotational diagram analysis ( \citealt{Turner1991,Sutton1995}) which assumes that the lines are optically thin and that the population levels are characterized by a single excitation temperature (LTE assumption). Integration of the transfer equation of the emission in a line with an upper energy level, E$_{u}$, leads to the expression (eg., \citealt{blake-1987, Araya2005}),
\begin{table*}
\begin{center}
\caption{Observed and derived parameters from CH$_{3}$CN observations.} \smallskip
\begin{tabular}{lcccccccc}\hline\hline
Core	&	\multicolumn{5}{c}{Velocity integrated flux density (Jy km s$^{-1}$)}	& $\theta$ 	&	T$_{rot}$	&	N$_t$(CH$_{3}$CN)	\\ \cline{2-6}
	&	$K = 0$	&	$K = 1$	&	$K = 2$	&	$K = 3$	&	$K = 4$	&	($\arcsec$)			&	(K)	&	(10$^{14}$ cm$^{-2}$)	\\ \hline
	\multicolumn{9}{c}{AGAL329} \\ \hline
mm-1	&	0.22	&	0.24	&	0.08	&	0.04	&	--	&	2.0	&	30$\pm$3	&	4.4$\pm$1.0 \\
mm-2	&	--	&	--	&	--	&	--	&	--	&	--	&	--		&	--	\\
mm-3	&	0.22	&	0.23	&	0.11	&	0.07	&	--	&	2.0	&	41$\pm$5	&	6.3$\pm$1.3 \\
mm-4	&	0.33	&	0.34	&	0.13	&	0.07	&	--	&	2.0	&	34$\pm$3	&	7.3$\pm$1.4 \\
mm-5	&	0.15	&	0.13	&	0.05	&	--	&	--	&	2.0	&	33$\pm$4	&	3.0$\pm$0.7 \\
mm-6	&	5.90	&	4.89	&	2.62	&	2.09	&	0.34	&	5.0	&	68$\pm$7	&	32.1$\pm$6.5 \\
mm-7	&	0.12	&	0.12	&	0.05	&	0.02	&	--	&	2.0	&	31$\pm$2	&	2.4$\pm$0.3 \\
mm-8	&	--	&	--	&	--	&	--	&	--	&	--	&	--	&	--	\\
mm-9	&	0.17	&	0.13	&	0.05	&	--	&	--	&	2.0	&	28$\pm$1	&	4.8$\pm$0.2 \\
mm-10	&	0.43	&	0.39	&	0.17	&	0.12	&	--	&	2.0	&	41$\pm$3	&	11.0$\pm$1.7\\\hline
	\multicolumn{9}{c}{AGAL333} \\ \hline
mm-2	&	0.08	&	0.09	&	0.03	&	--	&	--	&	2.5	&	31$\pm$12 &	1.1$\pm$0.7 \\
mm-4	&	0.07	&	0.05	&	0.02	&	--	&	--	&	2.0	&	30$\pm$1	&	0.51$\pm$0.03 \\
mm-5	&	0.04	&	0.02	&	0.01	&	--	&	--	&	3.0	&	24$\pm$1	&	0.50$\pm$0.06  \\
\hline
\end{tabular}
\end{center}
\label{tab:Trot}
\end{table*}
\begin{figure*}[ht!]
\centering
\includegraphics[trim={0.cm 0.cm 0.cm 0.cm}, clip, width=0.7\textwidth]{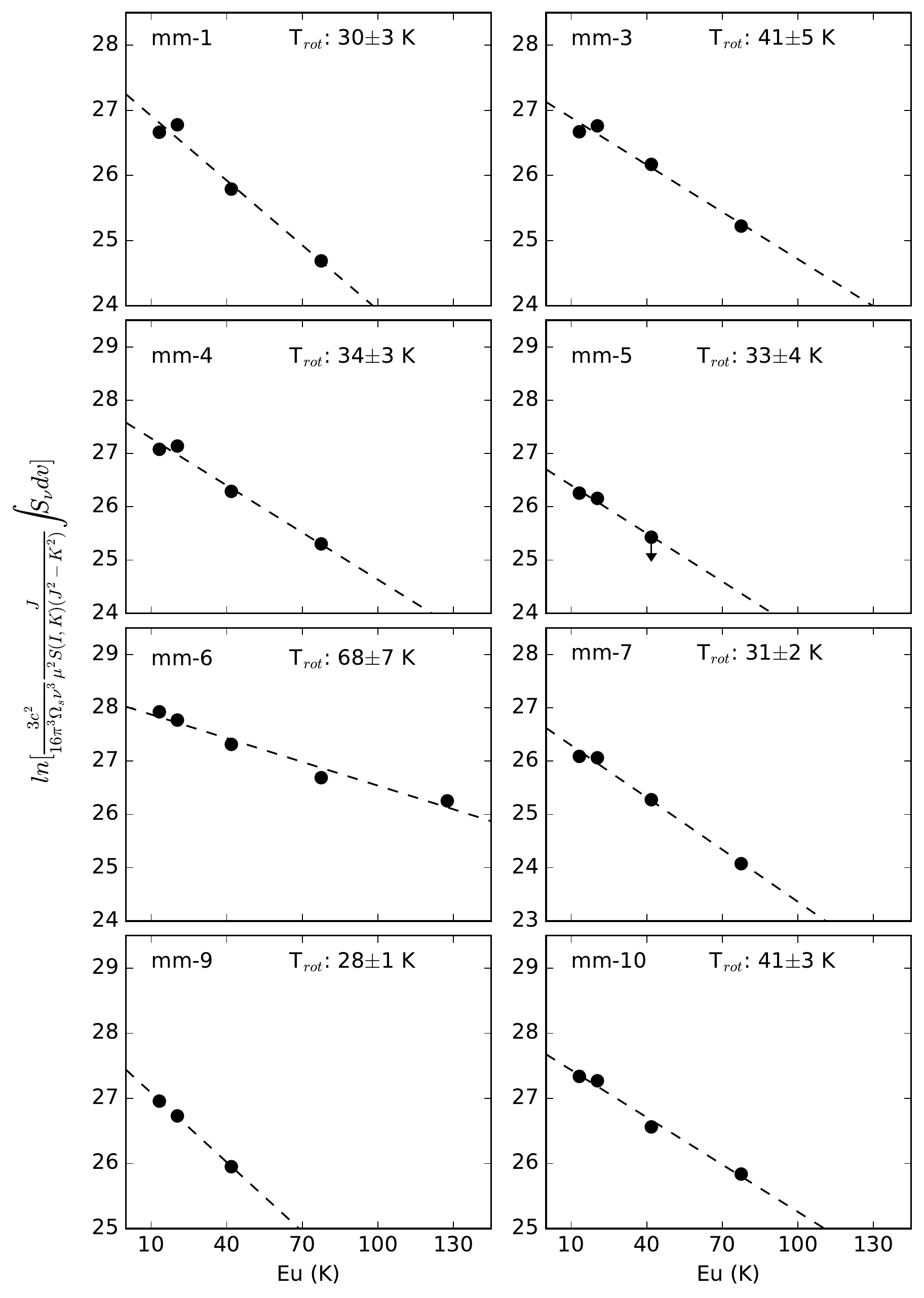}
\caption{CH$_{3}$CN rotational diagram for cores in the protostellar clump. The derived rotational  temperature 
is given in the upper right corner. \label{g329_trot_ch3cn}}
\end{figure*}
\begin{figure*}[ht!]
\centering
\includegraphics[trim={0.cm 0.cm 0.cm 0.cm}, clip, width=0.7\textwidth]{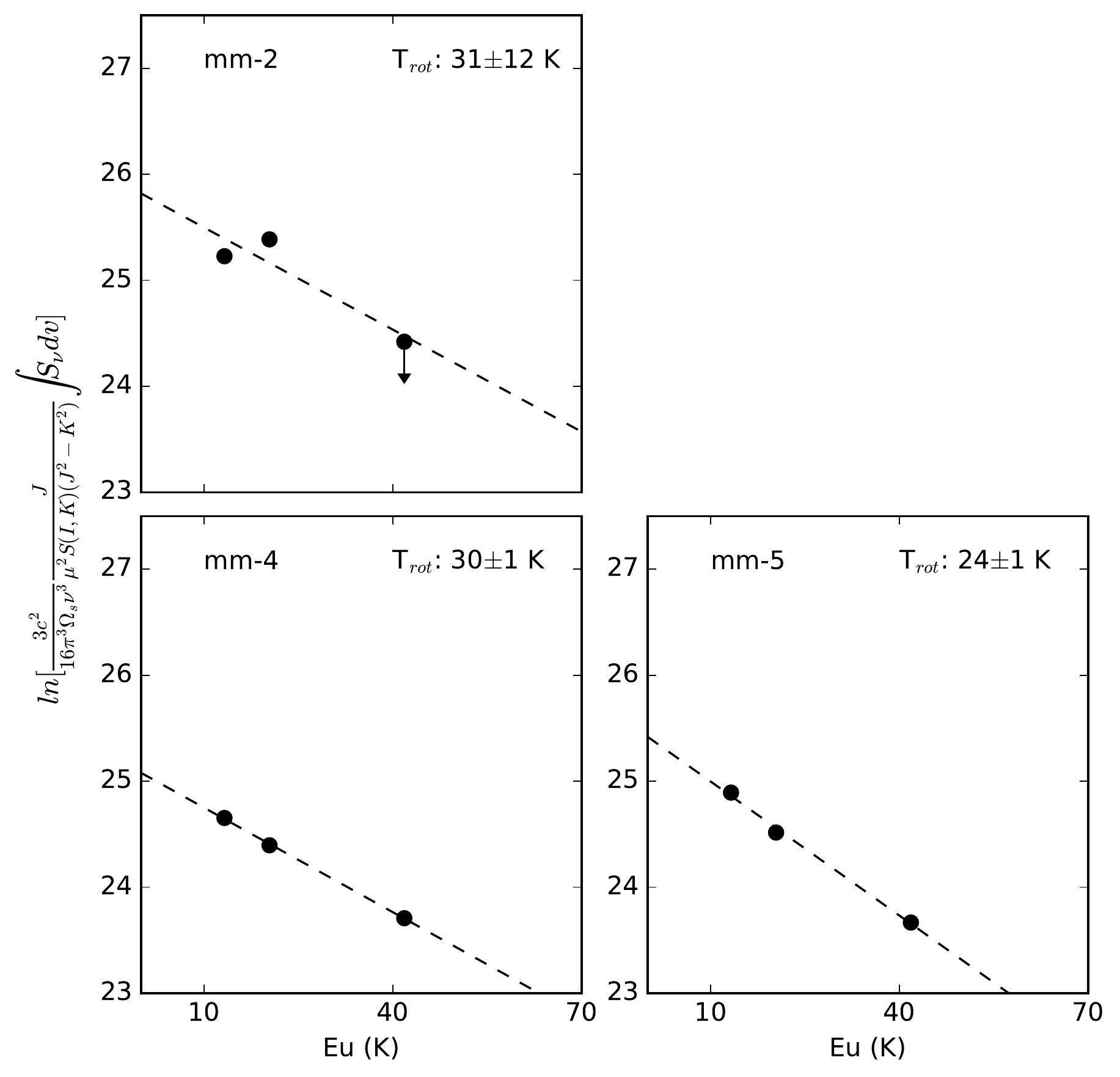}
\caption{CH$_{3}$CN rotational diagram for cores in the prestellar clump. The derived rotational  temperature 
is given in the upper right corner.\label{g333_trot_ch3cn}}
\end{figure*}
\begin{eqnarray*} \label{rotdiag}
ln\Big[\frac{3c^2}{16 \pi^3  \Omega_{s} \nu^3}\frac{J}{\mu^{2} S(I,K)(J^2 - K^2)}{\int{S_{\nu} dv}}\Big]=ln\Big[\frac{N_t}{Q(T_{rot})}\Big] - \Big[\frac{E_u}{kT_{rot}}\Big]~~,
\end{eqnarray*}
where $\int{S_{\nu} dv}$ is the velocity integrated flux density of the line, $\Omega_s$ the solid angle subtended by the source, $S(I,K)$  the degeneracy due to spin. 
For CH$_{3}$CN J = $5\rightarrow4$ transitions, the spin degeneracies S(I,K) are presented in Table \ref{table:ch3cnlinefreqs}}, $\nu$ and $\mu$  the transition frequency and dipole moment of the molecule,  respectively, T$_{rot}$ the rotational temperature, Q(T$_{rot}$) the rotational partition function, and N$_{t}$ is the total column density.

Rotational diagrams for AGAL329 and AGAL333 cores, for which at least three lines in the 5$_K$ - 4$_K$ K-ladder were detected above the 3$\sigma$ level, are shown in Figures \ref{g329_trot_ch3cn} and \ref{g333_trot_ch3cn}, respectively. The velocity integrated flux density, obtained by integrating the flux per beam over a circular region with angular radius given in col. 7 of Table \ref{tab:Trot}, are given in cols. 2 to 6 for K = 0, 1, 2, 3 and 4 lines, respectively.
From a least squares fit to the data we derived the rotational temperatures given in column 8 of Table \ref{tab:Trot}.   Clearly, the cores within the protostellar clump are  warmer than in the prestellar clump. The temperature of cores within the protostellar clump range from 28 to 68 K, with an average value of $\sim$38 K.   Within the prestellar clump only three cores (mm-2, mm-4 and mm-5) were detected in at least three 5$_K$ - 4$_K$ lines, for which we derived temperatures of 31, 30 and 24 K, respectively. 
Whether  the source of heating of the cores in the protostellar clump is due to the presence of embedded protostellar activity or from their gravitational collapse remains to be investigated.

\begin{table*}
\begin{center}
\caption{Observed and derived parameters from CH$_{3}$CN observations toward core mm-6 in AGAL329.}
\smallskip
\begin{tabular}{ccccccccc}\hline\hline
Region	&	\multicolumn{5}{c}{Velocity integrated flux density Jy (km s$^{-1}$)}	& 	$\Omega_{eff}$\tablenotemark{a}	&	T$_{rot}$	&	N$_t$(CH$_{3}$CN)	\\ \cline{2-6}
	&	$K = 0$	&	$K = 1$	&	$K = 2$	&	$K = 3$	&	$K = 4$		&	($\arcsec^2$)		&	(K)	&	(10$^{15}$ cm$^{-2}$)	\\ \hline
1	&	0.65	&	0.46	&	0.36	&	0.39	&	0.09	&	3.1		&	131$\pm$24	&	25.4$\pm$8.1 	\\
2	&	1.03	&	0.77	&	0.52	&	0.50	&	0.10	&	9.1		&	94$\pm$12	&	8.4$\pm$1.9 	\\
3	&	0.94	&	0.81	&	0.41	&	0.31	&	0.06	&	15.2		&	63$\pm$8		&	2.8$\pm$0.7  	\\
4	&	1.14	&	1.00	&	0.46	&	0.31	&	0.05	&	21.2		&	53$\pm$6		&	2.0$\pm$0.5 	\\
5	&	1.10	&	0.95	&	0.45	&	0.30	&	0.03	&	28.3		&	42$\pm$1		&	1.2$\pm$0.1 	\\
6	&	1.04	&	0.90	&	0.41	&	0.28	&	0.01	&	31.2		&	30$\pm$3		&	0.8 $\pm$0.2  	\\
\hline
\end{tabular}
\end{center}
\tablenotetext{a}{ Effective solid angle of the region.}
\label{tab:Trotmm6}
\end{table*}

The fit also gives the value of $\frac{N_t}{Q(T_{rot})}$ which allows to derive the CH$_{3}$CN column density. 
Using the following expression for the partition function of CH$_{3}$CN (\citealt{Araya2005}),
\begin{eqnarray*}
Q(T_{rot}) = 3.89 ~ \frac{T_{rot}^{1.5}}{(1-e^{-524.8/T_{rot}} )^2}~~,
\end{eqnarray*}
and the rotational temperatures of the cores
we derived the CH$_{3}$CN column densities given in column 9 of Table \ref{tab:Trot}. The cores in the protostellar clump have column densities ranging from 2.4x10$^{14}$~cm$^{-2}$ to 3.2x10$^{15}$~cm$^{-2}$, while the mm-2, mm-4 and mm-5 cores in the prestellar clump have column densities of 1.1x10$^{14}$, 5.1x10$^{13}$ and 5.0x10$^{13}$~cm$^{-2}$, respectively. 

It is worth to mention that one of the assumptions of the rotational diagram method applied above, namely that lines 
should be optically thin, it is well fulfilled by the emission in the J=5$_K\rightarrow4_K$ lines from all cores. For instance, for core mm-6 in AGAL329, the most extreme case since it exhibits the largest column density,  the optical depths in the  CH$_{3}$CN
$J=5_K\rightarrow4_K$, K = 0, 1, 2, 3, 4 lines are  0.23,  0.19, 0.11,  0.04 and  0.02, respectively.  If we use the more general rotational 
diagram method which takes into account optical depth effects (c.f., \citealt{Goldsmith1999}), the differences in the derived temperatures and column densities are well within the errors determined from the ``traditional" approach. 

\begin{figure*}
\centering
\includegraphics[trim={0cm 0.2cm 0cm 0.2cm}, clip,width=0.99\textwidth]{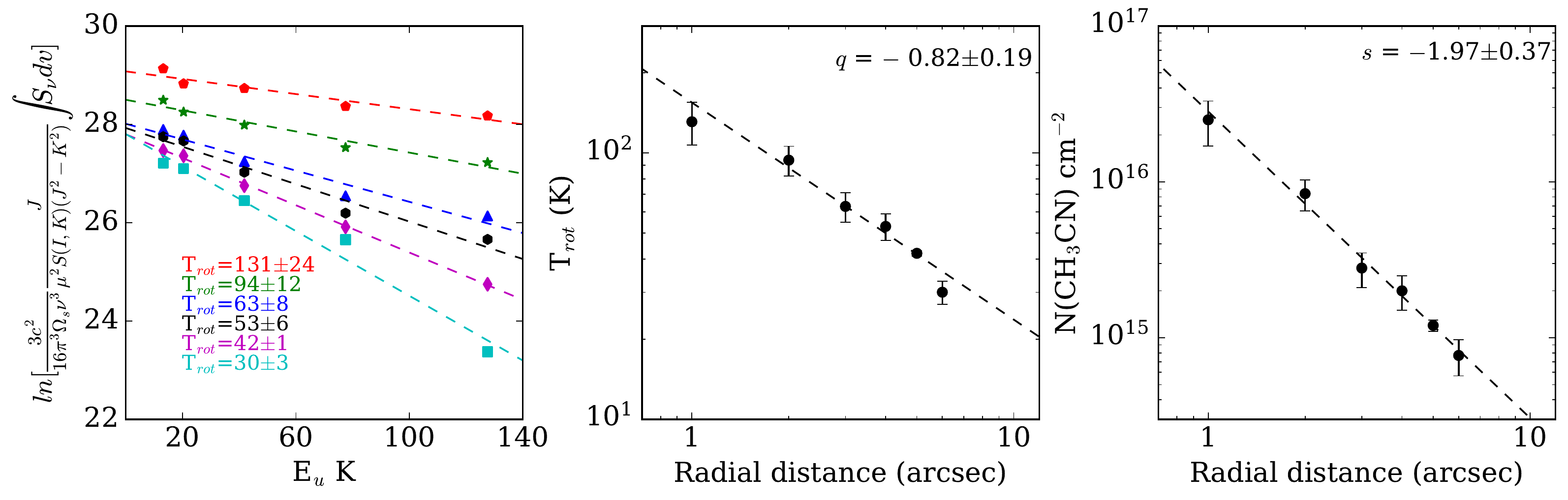}
\caption{Left: Rotational diagram of the CH$_{3}$CN emission from six different regions within core mm-6 in AGAL329 (see text for the description of the regions). The derived rotational temperatures are shown in the lower left corner. Middle: Rotational temperature dependence with radius.  Right: CH$_{3}$CN column density dependence with radius.  \label{trotprofile}}
\end{figure*}

Due to the large spatial extent of the CH$_{3}$CN emission from the central region of AGAL329, whose peak position coincides with the peak position of core mm-6,  it was possible to determine the dependence of the rotational temperature with radius.  Figure \ref{trotprofile} (left panel) shows rotational diagrams of the CH$_{3}$CN emission integrated over six different regions: an inner disk with a radius of 1$\arcsec$ and 5 circular annuli with inner radius from 1$\arcsec$ to 5$\arcsec$ 
and width of 1$\arcsec$. The derived rotational temperature and column densities are given in cols. 8 and 9 of Table \ref{tab:Trotmm6},
respectively. The rotational temperature decreases from 131$\pm$24 K at the peak position to 30$\pm$3 K at a radial distance of 6$\arcsec$ from the center. Also shown in the middle and right panels of Figure \ref{trotprofile} are, respectively,  the rotational temperature and CH$_{3}$CN column density dependence with radius. Power law fits to the rotational temperature profile
(T$_{rot}$ $\propto$ $r^{q}$) and  column density profile  (N $\propto$ $r^{s}$) give power law indices of $-0.8\pm0.2$ 
and  $-2.0\pm0.4$, respectively.

\subsubsection{Masses  \label{sec:mass}}
The mass of the cores were calculated from the continuum flux density, $S_{\nu}$,  using the expression,
\begin{equation}
M = \frac{S_{\nu} D^{2} R_{gd}}{ k_{\nu} B(T, \nu)} ~~,
\end{equation}
where $k_{\nu}$ is the dust mass absorption coefficient at frequency $\nu$,  B(T,$\nu$) is the Planck function at temperature T, D is the distance and $R_{gd}$ is the gas-to-dust ratio.  We assume $R_{gd}$=100 and $k_{100 GHz}$ = 0.21 cm$^{2}$ g$^{-1}$ corresponding to the dust grains with ice mantles at gas densities of 10$^{6}$ cm$^{-3}$ (\citealt{Ossenkopf1994}).  For the temperatures we used the values of the rotational temperatures derived from the CH$_{3}$CN observations (see \S \ref{ch3cn_temp}). For cores in which no rotational temperature is available we adopted the temperature of the clump. 
The masses, listed in col. 3 of Table \ref{coreparam}, range from 1.6 to 20 M${_\odot}$  for cores in the prestellar clump  and from 1.7  to 119 $M_{\odot}$ for cores in the protostellar clump. The uncertainties in the mass are estimated to be $\sim$40\%, considering errors in the flux density, temperature,  distance and dust mass absorption coefficient (c.f., \citealt{Sanhueza2017}). It is worth to note that to compute the mass of the central core in AGAL 329, we subtracted from the observed flux density at 100 GHz the expected contribution from free-free emission. The latter was determined from an extrapolation of the observed flux densities at 18 GHz and 22 GHz (\citealt{sanchez-monge2013}). The total mass in the form of cores 
is $\sim$6\% of the clump mass in the prestellar clump  and $\sim$23\% in the protostellar clump (see Table \ref{coremassdist}).

\subsubsection{Column densities \label{sec:colden}}

The source averaged H$_{2}$ column densities of the cores can be computed from the continuum flux density, $S_{\nu}$, using the expression,
\begin{equation}
N_{H_{2}} = \frac{S_{\nu}R_{gd}}{ \Omega_{c} \mu_{H_{2}} m_{H} k_{\nu} B(T, \nu) }  ~~,
\end{equation}
where $\mu_{H_{2}}$ = 2.8 is the molecular weight per hydrogen molecule, m$_{H}$ is the H-atom mass, $\Omega_{c}$ is the solid angle subtended by the core. 
Col. 4 of Table \ref{Table:coldensity} lists the source averaged column densities, computed using the flux density measured in circular regions with the angular radius given in col. 3 and as dust temperature the CH$_{3}$CN rotational temperature of the cores (or clump temperature for cores in which rotational temperature is not available).  They range from 6.0$\times 10^{22}$ to 2.6$\times 10^{23}$ cm$^{-2}$  for cores in the prestellar clump and from  1.0$\times 10^{23}$ to 7.5$\times 10^{23}$ cm$^{-2}$ for cores in the protostellar clump. 
The highest values of the H$_{2}$ column densities are found towards the centrally located cores, mm-6 in AGAL329 and mm-5 in AGAL333. The uncertainties in the column densities are estimated to be $\sim$35\%.

\begin{table*}[htb!]
\begin{center}
\caption{Source averaged H$_{2}$ and N$_{2}$H$^{+}$ column densities. \label{Table:coldensity}} 
\begin{tabular}{cccccc}\hline
Core & $S_{\nu}$ & $\theta$ & N(H$_{2}$) & N(N$_{2}$H$^{+}$) & X(N$_{2}$H$^{+}$) \\
	 &	(mJy)		&	($\arcsec$)	&	(10$^{23}$ cm$^{-2}$)		&	(10$^{14}$ cm$^{-2}$)	& (10$^{-9}$)		\\ \hline
\multicolumn{6}{c}{AGAL329}\\ \hline

mm-1 & 5.1 	& 2.5  & 3.9 & 16.1 & 4.2 \\
mm-2 & 0.9 	& 2.0 & 1.1 & 6.2 & 5.4 \\
mm-3 & 5.8 	& 3.0 & 2.2 & -- & -- \\
mm-4 & 10.7 	& 3.0 & 4.9 & -- & -- \\
mm-5 & 2.1 	& 3.0 & 1.0 & 8.6 & 8.5 \\
mm-6 & 93.8 	& 5.0 & 7.5 & 12.4 & 1.7 \\
mm-7 & 1.5 	& 2.0 & 1.8 & 2.1 & 1.2 \\
mm-8 & 3.3 	& 3.0 & 1.9 & 3.0 & 1.6 \\
mm-9 & 1.7 	& 1.5 & 3.9 & 3.6 & 0.9 \\
mm-10 & 8.4	 & 2.0 & 7.1 & 8.9 & 1.3 \\ \hline
Average &		&	& 3.5 	& 7.6 	& 3.1	\\ \hline
\multicolumn{6}{c}{AGAL333}\\ \hline
mm-1 & 0.7 	& 2.5 & 0.7 & 1.1 & 1.6 \\
mm-2 & 2.4 	& 3.0 & 1.2 & 3.0 & 2.5 \\
mm-3 & 1.4 	& 3.0 & 1.0 & 0.8 & 0.8 \\
mm-4 & 2.8 	& 3.0 & 1.5 & 8.3 & 5.5 \\
mm-5 & 3.8 	& 3.0 & 2.6 & 2.6 & 1.0 \\
mm-6 & 1.7 	& 2.5 & 1.8 & 1.4 & 0.8 \\
mm-7 & 1.3 	& 3.5 & 0.7 & 1.4 & 2.1 \\
mm-8 & 0.9 	& 2.0 & 1.4 & 3.0 & 2.1 \\
mm-9 & 0.7 	& 2.0 & 1.1 & 0.8 & 0.7 \\
mm-10 & 1.1 	& 3.5 & 0.6 & 0.9 & 1.6 \\\hline
Average &		&	& 1.3 	& 2.3 	& 1.9 	\\ \hline
\end{tabular}
\end{center}
\end{table*}

From the observations of the  N$_{2}$H$^{+}$ line emission it is possible to compute the source averaged column densities using the expression (eg., \citealt{Garden1991,Mangum2015}),

\begin{equation}
N_{tot}(N_{2}H^{+}) = \frac{3k}{8\pi^{3}\mu^{2}B} \frac{(T_{ex} + hB/3k)}{J_{u}} \frac{exp(\frac{E_{u}}{kT_{ex}})}{exp(h\nu/kT_{ex})-1} \int{\tau_{\nu} dv} ~~,
\end{equation}
where, E$_{u}$ is the upper level energy, B is the rotational constant of the molecule, T$_{ex}$ is the excitation temperature, $\nu$ is the frequency, $\mu$ is the dipole moment, Q$_{rot}$ is the partition function, $k$ is the Boltzmann constant, $h$ is the Planck constant and $\tau$ is the total optical depth. For observations of the J=1$\rightarrow$0 line, 
\begin{equation}
N_{tot}(N_{2}H^{+}) = 3.10\times10^{11} {(T_{ex} + 0.74)} \frac{exp(\frac{4.47}{T_{ex}})}{exp(\frac{4.47}{T_{ex}})-1} \int{\tau_{\nu} dv}\quad cm^{-2}  ~~,
\end{equation} 
where $dv$ is in km s$^{-1}$.

\begin{figure*}
\centering
\includegraphics[trim={0cm 0cm 0cm 0cm}, clip, width=0.6\textwidth]{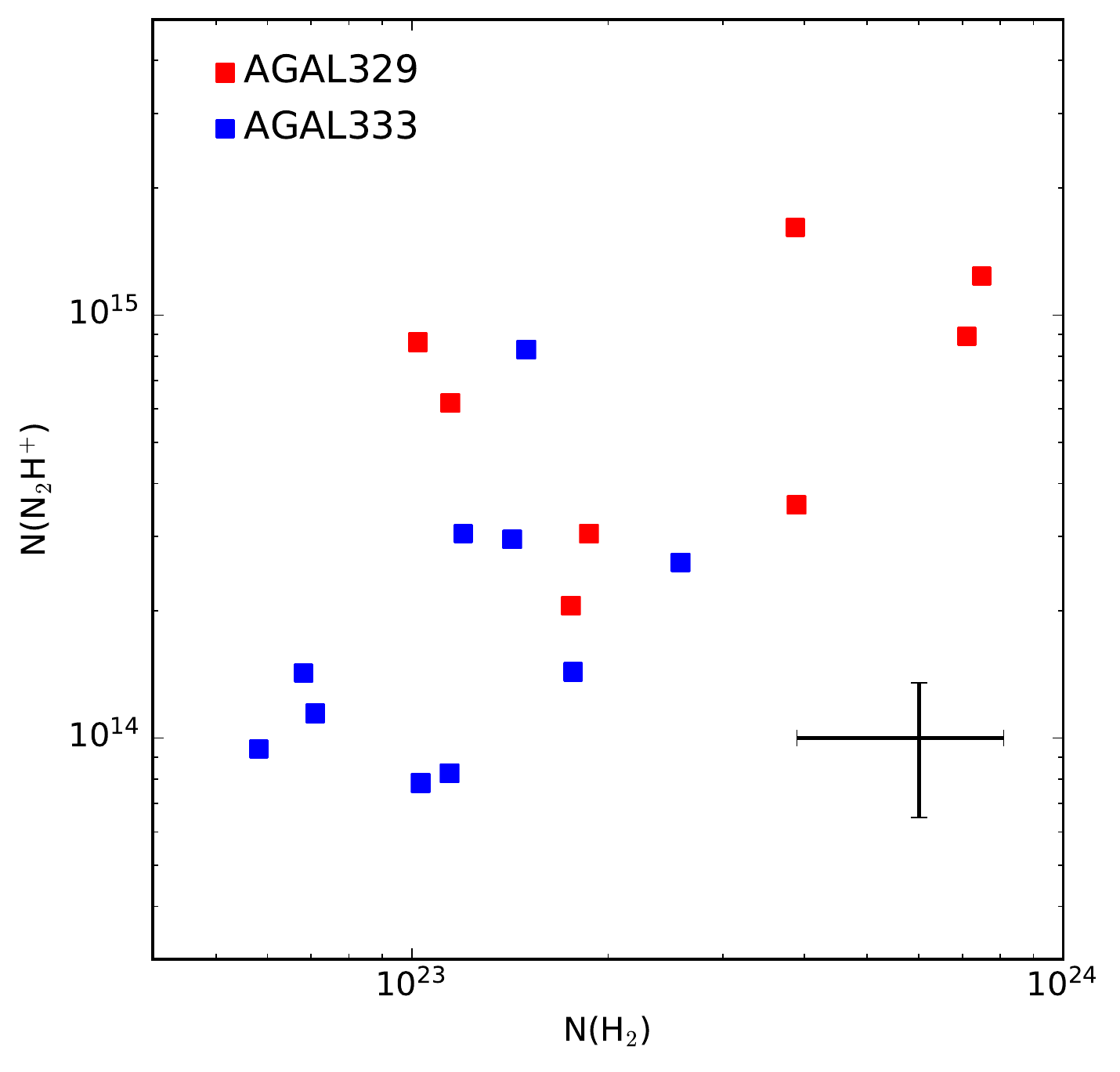}
\caption{H$_{2}$ column densities versus N$_{2}$H$^{+}$ column densities for prestellar cores (blue squares) and protostellar cores (red squares). Typical errors in column densities are shown in the lower right corner. \label{figure:colden_dist}}
\end{figure*}

Col. 5 of Table \ref{Table:coldensity} lists the N$_{2}$H$^{+}$ column densities of the cores 
computed from the above relation using the line widths and total optical depths determined from the HFS fit. 
For the temperature, we adopted the CH$_{3}$CN rotational temperature of the cores. For cores for which no rotational temperature is available the temperature of the clump was used. As shown in Figure \ref{figure:colden_dist}, which plots the H$_2$ versus N$_{2}$H$^{+}$ column densities, cores in the protostellar clump have typically larger H$_{2}$ and N$_{2}$H$^{+}$ column densities than cores in the prestellar clump. 
The average N$_{2}$H$^{+}$ column density of the cores in the prestellar and prostellar clumps are, respectively, 2.3$\times10^{14}$ cm$^{-2}$  and  7.6$\times10^{14}$ cm$^{-2}$ and the average H$_{2}$ column density are 1.3$\times10^{23}$ cm$^{-2}$ and  3.5$\times10^{23}$ cm$^{-2}$. The average abundance of N$_{2}$H$^{+}$ relative to H$_2$, computed as the ratio of the respective column densities, are 1.9$\times10^{-9}$ and 3.1$\times10^{-9}$ for cores in the prestellar and protostellar clumps, respectively, indicating that in protostellar cores the abundance of N$_{2}$H$^{+}$ is typically larger than in prestellar cores. An increase in the N$_{2}$H$^{+}$ abundance with evolutionary stage has also been reported for clumps (eg., \citealt{Sanhueza2012, Hoq-2013}).

\subsection{Mass distribution}

\begin{figure}[ht!]
\centering
\includegraphics[width=0.7\linewidth]{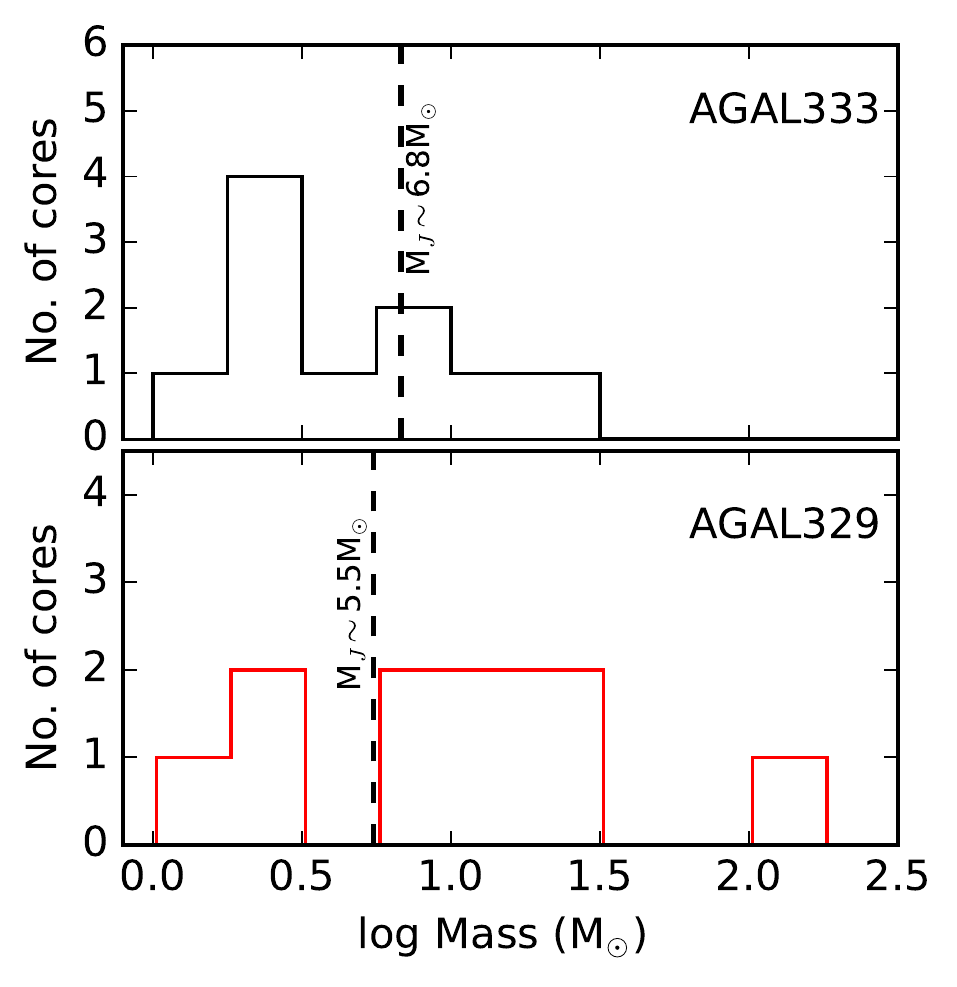}
\caption{Core mass distribution. Upper panel: AGAL333. Bottom panel: AGAL329.
The dotted line indicates the Jeans mass at the average conditions of the clump.\label{massdist} }
\end{figure}

Figure \ref{massdist} shows the distribution of the core masses in each clump. The dotted line indicates the value of the Jeans mass at the average clump conditions (see Table \ref{sampleparams}). The masses of the cores are of the order or slightly higher than the clump Jeans mass, except for the central core in AGAL329 which highly exceeds the Jeans mass. The number of cores detected in each clump ($\sim$10) is much smaller than the number of thermal Jeans masses contained in the clumps, of $\sim 160$, showing that fragmentation is not efficient during the early stages of evolution.  This conclusion was previously reported by \cite{Csengeri2017}, who found that the fragmentation of infrared quiet  MDCs at scales of  0.06 to 0.3 pc is limited, with most clumps hosting  typically 3 cores with masses of $\geq$ 40 M$_{\odot}$. 

Our observations with spatial resolution of $\sim$0.03 pc, ten times smaller than that of  \cite{Csengeri2017}, show that the number of cores per clump increases to 10, suggesting that we are resolving further fragmentation within MDCs. Recent studies of clumps with similar characteristics to those observed by \cite{Csengeri2017} have reported levels of fragmentation ranging from 5 to 20 cores when observed at scales of 0.03-0.05 pc (eg., \citealt{Lu2018, Contreras2018}).

\begin{figure}[ht!]
\centering
\includegraphics[ width=0.65\linewidth]{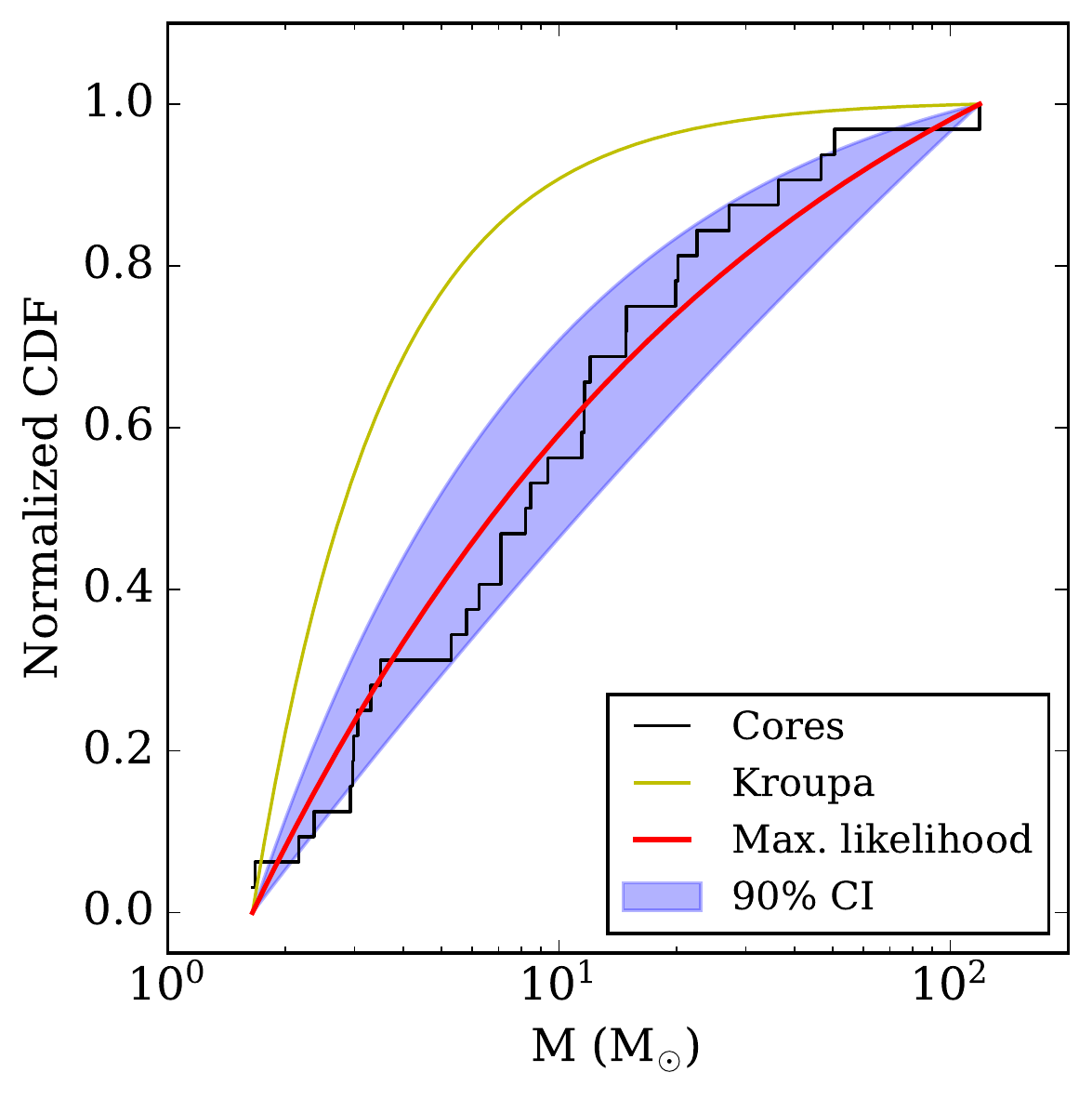}
\caption{The normalized cumulative distribution function (CDF) for the combined sample of cores within clumps AGAL333, AGAL329 and AGAL305. The red line and purple shaded area show the maximum likelihood estimation of the power law index fit and its 90\% confidence interval, respectively. The yellow line shows the initial mass function distribution from \citealt{Kroupa2001}.\label{masscmf} }
\end{figure}

\begin{deluxetable*}{lcc}
\tablecaption{Summary of core characteristics. \label{coremassdist}}
\tablehead{ \colhead{} &
\colhead{Prestellar clump } & \colhead{Protostellar clump }   \\  \colhead{} &
\colhead{AGAL333 } & \colhead{AGAL329}  }
\startdata
Number of cores         	     			& 	10          	&     10       				\\
Mean core size 					&	0.012 pc	&	0.014 pc	\\
Mean velocity dispersion				&	0.50 km s$^{-1}$	&	0.85 km s$^{-1}$	\\
$\Sigma$M$_{cores}$     				& 60 M$_{\odot}$ &     218 M$_{\odot}$        \\
$f$($\Sigma$M$_{cores}$/M$_{clump}$)	& 0.06		&	  0.23				\\
M$_{central core}$					& 20 M$_{\odot}$ & 	  119 M$_{\odot}$  		\\
$f$(M$_{central core}$/$\Sigma$M$_{cores}$)	& 0.33 	& 	0.55	\\
$f$(M$_{central core}$/M$_{clump}$)  	&	0.02		&	0.13	\\
	\enddata
\end{deluxetable*}

Figure \ref{masscmf} shows the normalized cumulative distribution function (CDF), also known as empirical cumulative distribution function (eCDF). We prefer to use the eCDF over the differential form of the core mass function because the later approach contains the numerical bias introduced by binning. Given the relatively small number of cores detected in each clump, we considered the combined sample of cores in the prestellar and protostellar clumps including, in addition, the cores detected by \cite{Servajean2019} towards the prestellar clump (AGAL305), which have similar characteristics to the clumps considered here. 
We adopt here as mass sensitivity limit the mass computed using a flux density equal to the 3$\sigma$ noise level and the temperature of the clump. The mass sensitivity limits  are  0.9, 0.7 and 1.4 M$_{\odot}$ for  AGAL329, AGAL333 and AGAL305, respectively.

Assuming that the core mass function (CMF) can be described by a power-law $dN/dM$ $\propto$ $M^{\alpha}$,
the value of the $\alpha$ index that best reproduces the eCDF, using the maximum likelihood estimator (MLE) method, is 
$-1.33\pm$0.15, and the 90\% confidence interval is $-1.58$ to $-1.08$.  This power-law index  is much shallower than that of the initial stellar mass function (IMF)
for stars with masses greater than 1M$_{\odot}$, of $-2.35$ (\citealt{Kroupa2001}), suggesting that in the early stage of fragmentation of clumps, high mass cores are more efficiently formed than low mass cores. Our result is alike to that of the recently reported ALMA study of CMF towards the HMSF region W43-MM1 (\citealt{MotteCmf2018}) and towards infrared dark clouds (IRDCs) (\citealt{liuCMF2018}) indicating the top-heavy nature of CMF in high-mass star forming regions.  

\subsection{Dynamical state}

To assess the dynamical state of  the cores we compute the virial parameter, $\alpha_{\rm vir}$, defined as $\alpha_{\rm vir} = M_{\rm vir}/M_{\rm dust}$, where M$_{\rm vir}$ is the virial mass defined as 
\begin{equation} M_{vir} = \frac{5\sigma^{2}R}{G} \end{equation}
where $\sigma = \sqrt{({\sigma_{th}^{2} + \sigma_{tur}^{2}})}$,  $\sigma_{th}$ and $\sigma_{tur}$ are the thermal and turbulent velocity dispersions, respectively, R is the radius and G the gravitational constant. The turbulent velocity dispersion was computed from the observed N$_{2}$H$^{+}$ or H$^{13}$CO$^{+}$ line widths and the thermal velocity dispersion was computed using the temperature given in col. 2 of Table \ref{coreparam} for a particle with a 
molecular weight of 2.33 (e.g., \citealt{Bertoldi-1992}). The virial mass and virial parameter of the cores are given in columns 8 and 9 of Table \ref{coreparam}, respectively. 
 Given the uncertainties in the values of the quantities that enter in the calculation of  $\alpha_{vir}$ we consider that cores which have $0.71 < \alpha_{vir} < 1.4$ are in virial equilibrium (i.e. we are considering an error of up to 40\%). 

In the prestellar clump, five cores are sub-virial (i.e. $\alpha_{vir} \leq 0.7$) indicating that their gravitational energy dominates their kinetic energy and, in absence of other means of support (e.g. magnetic energy), they are likely to be undergoing gravitational collapse. Two cores are in virial equilibrium and the three others have $\alpha_{vir} \ge 1.5$ suggesting that they may correspond to transient features. In the protostellar clump,  five cores are in virial equilibrium, two are sub-virial  and three have $\alpha_{vir} \geq 1.5$.

\subsection{The massive core at the center of the protostellar clump \label{collapsesignature}}

\begin{figure}
\centering
\includegraphics[width=0.6\linewidth]{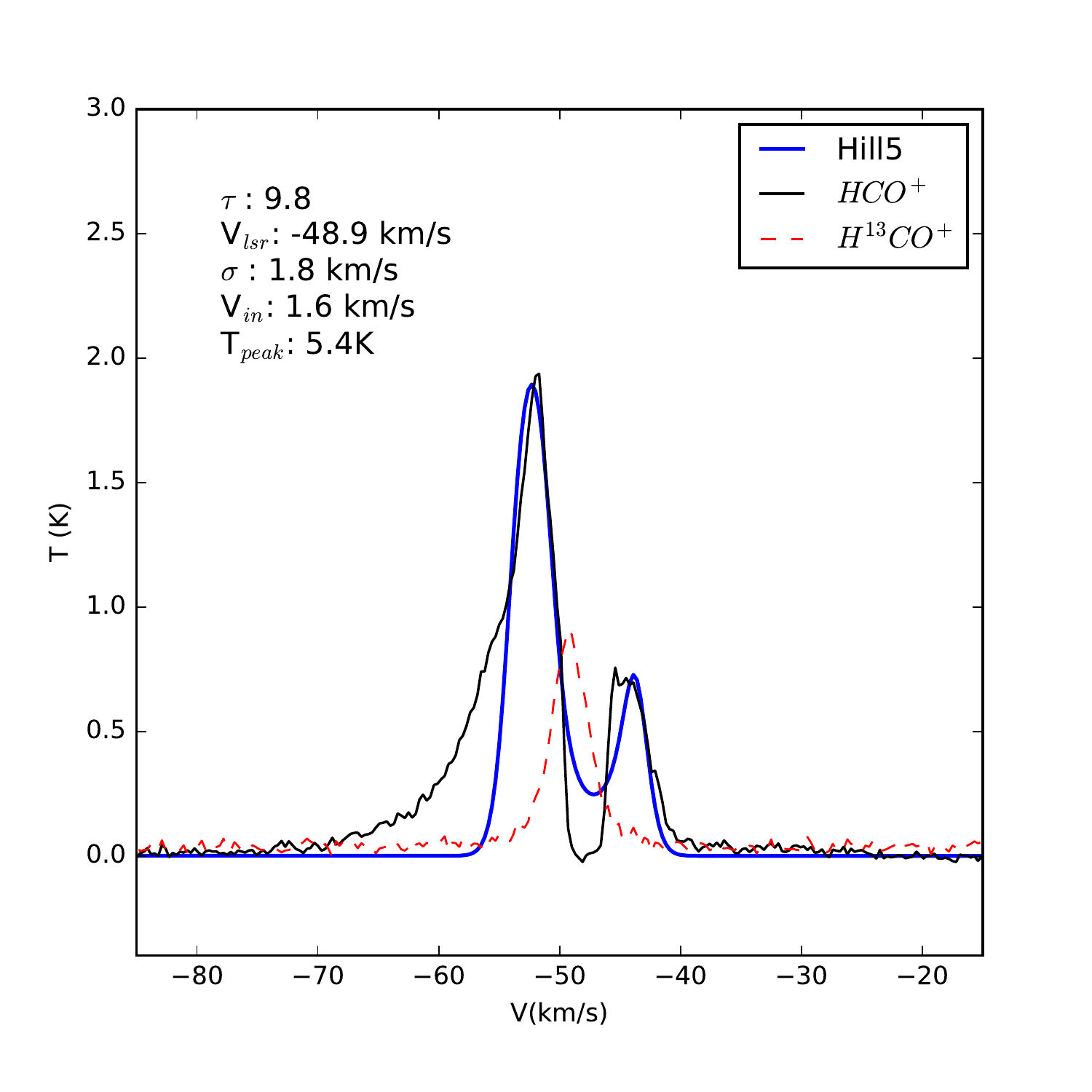}
\caption{Average spectra of the HCO$^{+}$ (black line) and H$^{13}$CO$^{+}$ (red line) emission from the central massive core in AGAL329. The blue line shows the best fit using the analytical infall models of \citealt{Devries2005}. Fitted parameters are given in the upper left corner. \label{hcop_model}}
\end{figure}

The massive (119 M$_{\odot}$) core located at the center of the protostellar clump has a virial parameter of 0.71, suggesting it is gravitationally bound and could be undergoing gravitational collapse. This hypothesis is strongly supported by the observed profiles in the optically thick HCO$^{+}$ line, which exhibits a double-peaked profile with the blue-shifted peak being brighter than the red-shifted peak, and in the optically thin H$^{13}$CO$^{+}$ line, which shows a single peak profile with a peak velocity near the dip in 
HCO$^{+}$ (see Figure \ref{hcop_model}). These characteristics of the line profiles are a classical signature of infalling motions (c.f. \citealt{mardones1997}).

To estimate the infall velocity we fitted the observed HCO$^{+}$ core spectrum with analytical infall models presented by \cite{Devries2005}. 
The best fit is attained with the ``Hill5" model (see Figure \ref{hcop_model}), which assumes that the excitation temperature increases inwards as a linear function of the optical depth, indicating an infall velocity, $v_{in}$,  of 1.6 km s$^{-1}$. We note that none of the simple models is able to reproduce the observed deep absorption feature (reaching zero intensity).  To reproduce it requires a more sophisticated modeling, which is beyond the scope of this work. From the derived values of the infall speed (1.6 km s$^{-1}$), core radius (0.03 pc), molecular weight $\mu_{H_2}$ (2.8), and molecular hydrogen density (1.5$\times10^{7}$cm$^{-3}$) we estimate a mass infall rate $\dot{M}$  (=4$\pi ~R^{2}~n(H_2) \mu_{H_2} m_H v_{in}$) of 1.9$\times$10$^{-2}$~M$_{\odot}$/yr,  value similar to those reported in other high-mass star forming regions (eg., \citealt{garay2002, beuther-2002,Contreras2018}).

\begin{figure}[!htb]
\centering
\includegraphics[width=0.65\linewidth]{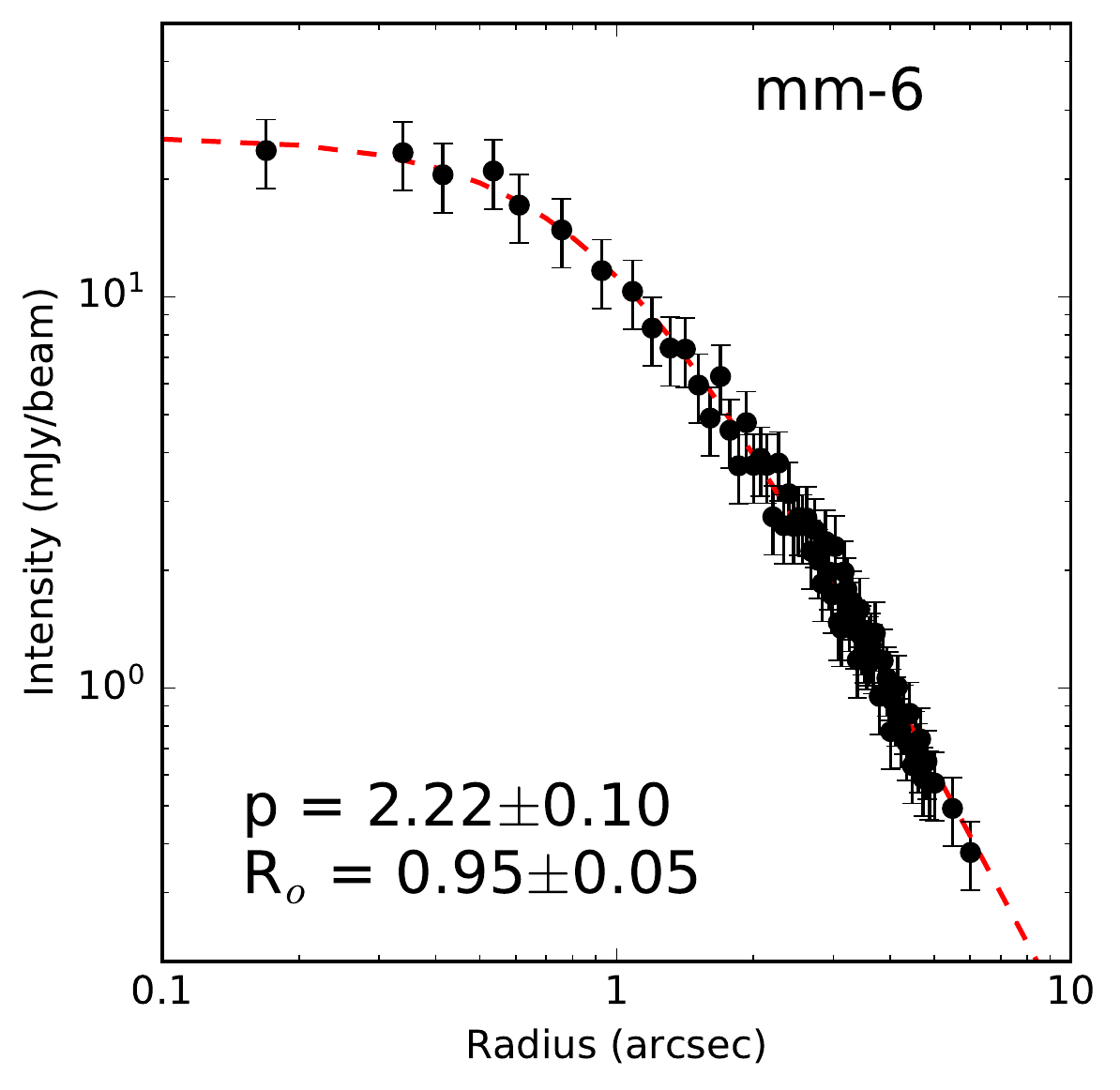}
\caption{Radial intensity profile of the massive core (mm-6) at the center of AGAL329.  The dotted red line correspond to a fit with a Plummer-like profile. Error bars correspond to 10\% errors in the observed intensities.\label{radial_profile}}
\end{figure}

The observed radial intensity profile of the massive core, shown in Figure \ref{radial_profile}, deviates significantly from a Gaussian profile but is well approximated with a Plummer-like radial intensity profile of the form, 
\begin{equation}
I(I_0,R_0,p)=\frac{I_0}{(1+(\frac{R}{R_0})^2)^\frac{p}{2}}
\end{equation}
where $p$ is the power law index, I$_0$ is the intensity at the Plummer radius R$_0$. The best fit (shown as a red line) indicates  a power law index $p$ of 2.2$\pm$0.1 and R$_0$ of 0.95$\pm$0.05\arcsec. \cite{Whitworth2001} have already shown that the density profile of cores undergoing collapse can be successfully fit with Plummer-like profiles.   

In summary, the observed and derived properties of this massive core, eg., collapse and outflow signatures, high mass infall rate, and Plummer-like density profile, are consistent with a picture in which large-scale collapse is feeding gas into this core which is forming a massive protostar at its center.

\subsection{Comparison with models of the fragmentation of clumps and the formation of high-mass stars} 

The formation of a cluster of stars is thought to proceed through a sequence of fragmentation, merging and collapse process within massive and dense clumps (eg., \citealt{Vazquez2009, Motte2018review}). Given the complexity of this process most of the recents advances in this field have been made through numerical simulations.  To better constrain the models, and hence to understand the formation of stars, it is crucial to know the initial conditions of the sequence. In particular, the properties of the cores at the early stages of evolution of MDCs are poorly known. 

Our ALMA observations of the two MDC in early stages of evolution, with spatial resolutions of $\sim$0.02 pc, allowed us to determine the characteristics of the fragmentation at early stages of evolution, such as the number of cores, their physical and kinematical characteristics and the initial core mass function (CMF). Both clumps have masses of $\sim 10^3$ M$_{\odot}$ and therefore can potentially form a cluster of stars, and in particular, from the empirical mass-size relationship (\citealt{Kauffmann2010}) will probably give rise to high-mass stars. Thus we can compare our findings with models of massive star and cluster formation, although we note that few of them have made predictions concerning the fragmentation at early stages of clump evolution. We recall that one clump is in the prestellar stage and the other in the protostellar stage, thus we can investigate differences in the cores  due to evolution.

The Competitive accretion model (\citealt{bonnell-2002,bonnel-2006}) proposes that a clump initially fragments into cores of thermal Jeans masses. These cores then accrete mass from the reservoir material in the clump via Bondi-Hoyle accretion. Cores located near the center of the gravitational potential accretes at a higher rate leading to the formation of a high mass protostar. In the competitive accretion scenario cores in an early stage after the fragmentation are expected to be subvirial. 

The turbulent core accretion model (\citealt{mckee-tan-2003}) proposes that stars form via a monolithic collapse of cores in virial equilibrium supported by the internal pressure due to turbulence and/or magnetic fields and hence should have masses much larger than the thermal Jeans mass. In this model the core mass distribution is then set at early evolutionary times, and therefore shall be similar to the initial stellar mass distribution (\citealt{Tan-2014}).  In addition, in this scenario cores are expected to be virialized  (\citealt{mckee-tan-2002,krumholz-bonnell-2009}).

We found that the number of cores detected in both clumps is considerably smaller than the number of thermal Jeans masses contained in the clumps (M/M$_J$ $\sim 160$) showing that fragmentation is not efficient during the early stages of evolution. In addition, the fraction of total core mass to clump mass is 6\% in the prestellar clump and rises to $\sim$23\% in the protostellar clump. Since the number of cores in both clumps is similar, and the fact that the masses of the cores in the protostellar clump are typicaly higher than the masses of the cores in the prestellar clump, which are of the order of the clump Jeans mass, supports the hypothesis of a continuous increase in core masses due to accretion from the prestellar to the protostellar stage.

A large fraction of the cores within the prestellar clump  (5 out of 10) are sub-virial ($\alpha < 0.7$), two are virialized ($0.7 <\alpha < 1.4$) and the remaining three (with $\alpha \geq$ 1.5) are most likely transient features. On the other hand, five out of ten cores in the protostellar clump are virialized and two are in sub-virial states. 

These results support the view of a globally collapsing turbulent clump undergoing gravitational fragmentation.  In this scenario,
during the early stages of evolution (AGAL333 clump) most of the formed cores should have masses typical of the thermal Jeans mass and be in sub-virial states. In a more advance stage (AGAL329 clump), the gas is funneled down to the center of the potential 
and the centrally located core continue to accrete gas at a high rate, becoming the most massive one.

\section{Summary \label{sec:s6}}

We carried out ALMA band 3 observations of 3mm dust continuum and molecular emission, in lines of HCO$^{+}$, H$^{13}$CO$^{+}$, N$_{2}$H$^{+}$ and CH$_{3}$CN, towards two massive and dense clumps in early, but distinct, stages of evolution, one in a prestellar stage (AGAL333.014-0.521) and the other in a protostellar stage (AGAL329.184-0.314). The goal was to reveal the physical and dynamical characteristics of the small-scale structures (or cores) within these clumps.  The results are summarized as follows.

1) From the 3mm continuum images we identified,  using the Clumpfind and Dendogram algorithms, about 10 cores within each clump. The cores in the prestellar clump, which are mainly distributed in a long filamentary structure running from NE to SW across  the clump, have dust derived masses from 1.6 to 20  M$_{\odot}$, sizes from 0.006 to 0.022 pc (1200 to 4500 AU) and densities from 3.0$\times$10$^6$ to 3.5 $\times$10$^7$ cm$^{-3}$. The cores in the protostellar clump have dust derived masses from 1.7 to 119 M$_{\odot}$, sizes from 0.005 to 0.030 pc (1000 to 6200 AU ) and  densities from 4.0$\times$10$^6$ to 5.3 $\times$10$^7$ cm$^{-3}$. The fraction of total core mass relative to the clump mass is $\sim$6\% in the prestellar clump and $\sim$23\% in the protostellar clump.  Most cores in the prestellar clump have masses within a factor of a few from the Jeans mass of the clump. However, the total number of cores is significanty  smaller than the number of Jeans masses in the clump indicating that fragmentation is inefficient during the early stages of evolution of clumps. 

2) Molecular emission was detected towards both clumps in all four observed species.  Of these, the N$_{2}$H$^+$ emission is the brightest and most extended one and the one that best correlates with the continuum emission morphology. 

{\it Prestellar clump.} The morphologies of the N$_{2}$H$^{+}$ and H$^{13}$CO$^{+}$ emission from AGAL333 are similar, delineating a complex network of filamentary structures across the whole region.  
The velocity field of the N$_{2}$H$^+$ emission shows a significant velocity gradient, of 4.7 km s$^{-1}$  pc$^{-1}$, in a NE to SW direction, across the whole clump. The mass required to explain this as due to gravitationally bound rotation is  460 M$_\odot$ within a radius of 0.45 pc.  CH$_{3}$CN emission is only detected towards the NE region of this clump.
 
{\it Protostellar clump.} The morphologies of the line emission from AGAL329 are noticeably different in the four observed transitions.
The N$_{2}$H$^{+}$ emission arises from a bright central region, with three distinct condensations, and an extended envelope of  emission which is highly correlated with the absorption feature seen in the 8 $\mu$m Spitzer image. 
The HCO$^{+}$ emission towards the central region, shows a banana-like morphology which is roughly coincident with the two westernmost N$_{2}$H$^{+}$ condensations, but no HCO$^{+}$ emission is seen from the eastern N$_{2}$H$^{+}$ condensation. In addition, the HCO$^{+}$ image shows two conspicuous features: a bright clumpy structure, located $\sim$18$\arcsec$ south of the central region, elongated in the NE-SW direction, (the South feature),  and a weak V shaped feature located $\sim10\arcsec$  west from the central region (the West feature).
The most prominent features in CH$_{3}$CN are a bright central region whose peak position coincides with the peak position of 
the central core mm-6, and a  bright V shaped region coincident with the West feature seen in HCO$^{+}$.

3) Emission in the N$_{2}$H$^{+}$, HCO$^{+}$ and  H$^{13}$CO$^{+}$ lines was detected towards all continuum cores within AGAL329 and all cores, except mm-9, in AGAL333. The line widths, determined from observations of the optically thin H$^{13}$CO$^{+}$ and N$_{2}$H$^{+}$  lines, are usually smaller for cores within the prestellar clump than for cores within the protostellar clump. The average line widths  are 1.2 km s$^{-1}$ and 2.0 km s$^{-1}$ for  AGAL333 and AGAL329 cores, respectively. The explanation for  the large line widths in cores within the protostellar clump is not straightforward, it may reflect an increase in the level of turbulence due to the beginning of star formation activity,  magnetic fields, or an increase in the gas velocities due to collapse motions.
Emission in CH$_{3}$CN was detected from all continuum cores in the protostellar clump and only weak emission was detected towards five cores in the prestellar clump.  The temperatures, derived from a rotational diagram analysis of the emission in the CH$_{3}$CN J=5$_K$ - 4$_K$ lines, range from 28 to 68 K for cores in protostellar clump and from 24 to 31 K for cores in the prestellar clump. 

4) Five cores within AGAL333 exhibit inverse P-Cygni profiles in the HCO$^{+}$ line and five are sub-virial (virial parameters smaller than 0.7), indicating that most cores within the prestellar clump seem to be undergoing contracting motions. Within AGAL329, most cores exhibit double peak profiles in HCO$^{+}$, with a stronger blueshifted peak, and single peaked profiles 
in H$^{13}$CO$^{+}$,  with a velocity in between those of the HCO$^{+}$ peaks. 
These pair of profiles in optically thick (HCO$^{+}$) and optically thin (H$^{13}$CO$^{+}$) lines are signature of infalling gas, which we suggest is associated to the global collapse of the protostellar clump. 
 
5) The core at the center of the protostellar clump is the most massive one, has a Plummer-like intensity profile and exhibits line profiles which are characteristic of infalling gas,  indicating an infall velocity of 1.6 km~s$^{-1}$ and a mass infall rate of of 1.9$\times$10$^{-2}$~M$_{\odot}$/yr. These results  convey the idea that this core is still acquiring mass via the gravitational focusing of gas from a globally collapsing clump towards its large potential well. In addition, this core is associated with a bipolar outflow in HCO$^{+}$ possibly driven by a recently formed high-mass protostar at its center. 
 
6) The core mass function for the combined sample of cores within three clumps (AGAL329, AGAL333  and AGAL305 from \citealt{Servajean2019}) has a power law index of $-1.33\pm$0.15, much flatter than that of the IMF for stars with masses greater than 1
M$_{\odot}$.  The top-heavy nature of the CMF suggests that at an early stage of clump fragmentation, high mass cores are more efficiently formed than low mass cores.

In summary, we conclude that we are witnessing the evolution of the dense gas in globally collapsing MDCs,  with AGAL333 showing  the initial stage of fragmentation, producing cores that are individually collapsing, while in AGAL329 we are seeing a later stage in which a considerable fraction of the gas has been gravitationally focused into the central region.

\acknowledgments{G.G. and S.N. gratefully acknowledge support from CONICYT projects PFB-06 and  AFB170002. This paper makes use of the following ALMA data: ADS/JAO.ALMA\#2016.1.00645.S. ALMA is a partnership of ESO (representing its member states), NSF (USA) and NINS (Japan), together with NRC (Canada), MOST and ASIAA (Taiwan), and KASI (Republic of Korea), in cooperation with the Republic of Chile. The Joint ALMA Observatory is operated by ESO, AUI/NRAO and NAOJ. 

}

\software{CASA (\citealt{casa2007}), GILDAS/CLASS (\citealt{pety2005gildas,Gildas2013}), Matplotlib (\citealt{matplotlib}), SciPy (\citealt{scipy2019}), Astropy (\citealt{astropy:2013}), Pyspekit (\citealt{Pyspekit})}

\bibliographystyle{aasjournal}
\bibliography{bibliography}

\end{document}